\documentclass[preprint]{aastex6}
\usepackage{graphicx}

\newcommand{\revis}[1]{{ #1}}

\newcommand{\be}{\begin{equation}}
\newcommand{\bel}[1]{\begin{equation}\label{eq:#1}}
\newcommand{\ee}{\end{equation}}
\newcommand{\bd}{\begin{displaymath}} 
\newcommand{\ed}{\end{displaymath}}   
\newcommand{\bea}{\begin{eqnarray}}
\newcommand{\beal}[1]{\begin{eqnarray}\label{eq:#1}}
\newcommand{\eea}{\end{eqnarray}}

\newcommand{\eqref}[1]{\ref{eq:#1}}

\newcommand{\lsim }{{\lower0.8ex\hbox{$\buildrel <\over\sim$}}}
\newcommand{\gsim }{{\lower0.8ex\hbox{$\buildrel >\over\sim$}}}

\def\aap{ A\&A}

\def\apjs{ApJ Supp}

\def\simge{\mathrel{%
   \rlap{\raise 0.511ex \hbox{$>$}}{\lower 0.511ex \hbox{$\sim$}}}}
\def\simle{\mathrel{
   \rlap{\raise 0.511ex \hbox{$<$}}{\lower 0.511ex \hbox{$\sim$}}}}

\newcommand{\Msun}{\ifmmode {M_{\odot}}\else${M_{\odot}}$\fi}
\newcommand{\Lsun}{\ifmmode {L_{\odot}}\else${L_{\odot}}$\fi}
\newcommand{\Rsun}{\ifmmode {R_{\odot}}\else${R_{\odot}}$\fi}

\shorttitle{Turbulent Molecular Clouds Regulated by Radiation Feedback}
\shortauthors{Raskutti, Ostriker, \& Skinner}

\begin{document}
\title{Numerical Simulations of Turbulent Molecular Clouds Regulated by
  Radiation Feedback Forces II: Radiation-Gas Interactions and Outflows}  

\author{Sudhir Raskutti}
\author{Eve C. Ostriker}
\author{M. Aaron Skinner}
\affiliation{Department of Astrophysical Sciences, Princeton University, Princeton, NJ 08544, USA}

\begin{abstract}
Momentum deposition by radiation pressure from young, massive stars
may help to destroy molecular clouds and unbind stellar clusters by
driving large-scale outflows.  We extend our previous numerical
radiation hydrodynamic study of turbulent, star-forming clouds to
analyze the detailed interaction between non-ionizing UV radiation and
the cloud material.  Our simulations trace the evolution of gas and
star particles through self-gravitating collapse, star formation, and
cloud destruction via radiation-driven outflows.  {These models are
  idealized in that we include only radiation feedback and adopt an
  isothermal equation of state.}  Turbulence creates a structure of
dense filaments and large holes through which radiation escapes, such
that only $\sim 50\%$ of the radiation is (cumulatively) absorbed by
the end of star formation.  The surface density distribution of gas by
mass as seen by the central cluster is roughly lognormal with
$\sigma_{\rm ln \Sigma} = 1.3-1.7$, similar to the {
  externally-projected surface} density distribution.  This allows low
surface density regions to be driven outwards to nearly $10$ times
their initial escape speed $v_{\rm esc}$. Although the velocity
distribution of outflows is broadened by the lognormal surface density
distribution, the overall efficiency of momentum injection to the gas
cloud is reduced because much of the radiation escapes.  The mean
outflow velocity is approximately twice the escape speed from the
initial cloud radius.  Our results are also informative for {
  understanding galactic-scale wind driving by radiation, in
  particular the relationship between velocity and surface density for
  individual outflow structures, and the resulting velocity and mass
  distributions arising from turbulent sources}.
\end{abstract}


\maketitle

\section{Introduction}
\label{Sec:Introduction}

Momentum deposition in Giant Molecular Clouds (GMCs) by radiation from
young O stars is often proposed as a means for driving turbulence, propelling 
high velocity outflows \citep{
  Matzner2002,MatznerJumper2015,Goldbaum2011,
  Hopkins2011, Hopkins2012,
  KrumholzThompson2012,  ThompsonKrumholz2016,  Thompson2015}, 
and eventually causing cloud destruction \citep{Odell1967,
  Elmegreen1983,Scoville2001, KrumholzMatzner2009, Fall2010,
  Murray2010, KrumholzDekel2010, Sales2014, SkinnerOstriker2015, Kim2016,
Raskutti2016}. In extreme galactic environments, radiation pressure may play an 
important role in helping to support the overall ISM against gravity
(limiting further star formation) and/or launching galactic-scale
winds \citep{Harwit1962,Scoville2003,Thompson2005,
  Murray2005,Hopkins2011,Murray2011, OstrikerShetty2011, Hopkins2012,
  ZhangThompson2012}.  However, the integrated mass and momentum in
these outflows, their detailed density and velocity statistics, and
the lifetime star-forming efficiency and destruction timescale of GMCs
all depend strongly on how radiation interacts with the highly
turbulent and inhomogeneous gas and dust within these clouds.

One of the hallmarks of GMC structure is the clumpy, filamentary morphology  
produced by the interaction between gravitational collapse and
turbulence
\citep[e.g.,][]{McKeeOstriker2007,Andre2014}. The initially turbulent
cloud is characterized by a lognormal column density PDF
\citep{Ostriker2001,VazquezSemadeni2001, Padoan2004a, Padoan2004b,
  Goodman2009, Kainulainen2009, Lombardi2010,Brunt2010, Price2011,Butler2014,
   Schneider2013, Schneider2015,Lim2016}, with most of the mass
concentrated well above the mean (area-weighted)
cloud surface density. Gravitational
collapse further enhances this contrast, and the PDF may develop a
power-law tail at the high column densities associated with star-forming 
cores \citep[][]{Klessen2000, Vazquez-Semadeni2008,
  Federrath2008, Kritsuk2011, BallesterosParedes2011, Collins2012,
  FederrathKlessen2013, Lee2014,Lombardi2015}.  It is therefore clearly necessary
to understand how radiation from dispersed stellar sources is filtered
through such a top-heavy density PDF in order to understand
radiation-driven outflows.

Simultaneously capturing the gravito-turbulent density structure while
accurately modelling momentum injection and outflows driven by star formation
feedback is very much a
work in progress. Instead, almost all semi-analytic studies and many numerical
ones explicitly (or implicitly) assume spherical symmetry
\citep{Matzner2002, KrumholzMatzner2009, Murray2010, Murray2011,
  DekelKrumholz2013, Sales2014,Kim2016}, so that the evolution is
set by the ratio of total momentum injection to gravity.  When both
radiation and gravity are produced by a central point source, their
relative strength may be captured through an average Eddington factor, 
$f_{\rm Edd}$.
Under spherical symmetry, once a shell becomes super-Eddington, i.e., 
$f_{\rm Edd} \gtrsim 1$, it can be driven outwards to asymptotic
velocities exceeding the escape speed of the star-forming region.
\citet{Murray2010,  Murray2011} and  \citet{Thompson2015} have argued that 
$\sim 10^4$~K gas 
often seen emerging from starburst galaxies at velocities between tens
and hundreds of kilometers per second \citep{Heckman1990, Steidel1996,
  Franx1997, Pettini2000, Pettini2001, Shapley2003, Martin2005,
  Rupke2005, Tremonti2007, Weiner2009, Menard2009}
may have been accelerated by radiation forces from massive central star
clusters.  
However, it remains unclear which theoretical results based on spherical
symmetry carry over and which must be strongly modified when the
filamentary gas structure is taken into account. \citet{Murray2010},
\citet{Hopkins2011}, and \citet{ThompsonKrumholz2016}
considered parameterizations of the column density PDF as a lognormal
distribution, with the first two studies concluding that the
mean radiation force
would not differ substantially from the case of a uniform gas distribution,
and the last pointing out that some lines of sight could be super-Eddington
(enabling wind driving) even if the average conditions are sub-Eddington.
{Because the variance in the lognormal increases with turbulent Mach
  number, the radiation-matter coupling is expected to be much more affected by
  inhomogeneity in the cold ISM phase than the warm ISM phase, and this
could also have implications for driving cold vs. warm galactic winds.}

Numerical simulations of turbulent, self-gravitating clouds have
tended to concentrate on the dynamics of expanding HII regions driven
by ionized gas pressure, which has generally been found to be inefficient
at producing outflows and dispersing clouds \citep{
  Dale2005, Dale2012,Dale2013a,
  Vazquez-Semadeni2010, Colin2013, Walch2012}. A series of
simulations by \cite{Dale2005, Dale2012, Dale2013a} found that
filamentary structure reduces the fraction of gas that is ionized
to a small portion of the cloud mass, with ionized gas
escaping to low density regions. As shown by these studies, pressure from photoionized
gas was capable of expelling substantial material only when the escape
speed was low compared to the ionized gas sound speed ($\sim 10~{\rm
  km~s^{-1}}$).

Overall, radiation forces from the non-ionizing component are expected
to dominate over ionized gas pressure in clouds with high surface density
and cloud mass 
\citep{Fall2010,Kim2016}, and there is some observational
support for this \citep{Lopez2011, Lopez2014, Pellegrini2007,
  Pellegrini2010}. Over the lifetime of a star cluster,
the influence of radiation pressure will still be
significantly less than that of Type II supernovae (SNe)
\citep{MacLowKlessen2004, ElmegreenScalo2004}, which inject an order
of magnitude more momentum to the ISM overall
(\citealt{OstrikerShetty2011}; see also
\citealt{KimOstriker2015, IffrigHennebelle2015, WalchNaab2015,
  Martizzi2015} for assessments of momentum injection by
SNe). However, the delay of SNe by $3-30~{\rm Myr}$ means that the
filamentary environment into which they deposit their energy will
likely already have been shaped by other forms of
feedback. \cite{WalchNaab2015} find that clouds that have
previously been photoionized absorb 50\% more SN energy and are
more likely to be disrupted. Meanwhile, \cite{IffrigHennebelle2015}
found that supernovae exploding near the edge of a cloud have less impact
overall, especially on transferring momentum to the dense gas,
than fully embedded supernovae; this again suggests that preprocessing of GMCs
by radiation will be important to the subsequent interaction with supernova
blast waves.  Thus, the significant radiation pressure present in massive clouds and clusters may 
play as crucial a role in the dispersal of GMCs 
even though supernovae are intrinsically more powerful.  

To date only a few simulations have addressed the effect of either
direct or reprocessed radiation on turbulent molecular gas via fully self-consistent 
radiation hydrodynamic (RHD) simulations \citep{KrumholzThompson2012,
  KrumholzThompson2013, Davis2014, SkinnerOstriker2015,
  Raskutti2016}. These have generally found that momentum deposition
is less efficient in a turbulent, filamentary medium than for the case of a
uniform, spherically symmetric shell.  For plane-parallel simulations
of IR radiation interacting with gas using opacity $\kappa \propto
T^2$ with a fixed external gravitational potential and fixed total
flux \citep{KrumholzThompson2012,Davis2014}, the measured radiation
force was reduced by tens of percent up to a factor $\sim 2$ due to
the anticorrelation of matter and radiation in a turbulent medium.
The simulations of \citet{SkinnerOstriker2015}, with fixed opacity, and
including self gravity from both gas and the star particles that form over time,
found similar reductions in the radiation force due to matter/radiation
anticorrelations.  

This is the third paper in which we apply the \textit{Hyperion} RHD
code \citep{SkinnerOstriker2013} to consider evolution of
self-gravitating, turbulent clouds in response to the radiation formed
by stars in collapsed regions.  \citet{SkinnerOstriker2015} considered
the regime of massive clouds that are optically thick to reprocessed
radiation. That work showed that reprocessed radiation only expels
mass when the Eddington ratio for IR radiation, $f_{\rm Edd,*} \equiv
\kappa_{\rm IR}\Psi / (4 \pi G c)$, exceeds unity, where $\kappa_{\rm
  IR}$ is the mean IR opacity and $\Psi$ the mean light-to-mass ratio
of stars.  However, even for cases with $f_{\rm Edd,*}>1$, the net star-forming
efficiency does not follow the scaling prediction 
($\varepsilon_{\rm final} \propto \kappa_{\rm IR}^{-1}$) from simple spherical
shell models, because radiation and matter tend to be more
anticorrelated at higher $\kappa_{\rm IR}$.  For clouds destroyed by
radiation, the mean velocity of ejected gas is $\sim 1-2$ times the
cloud's escape speed at its surface.

In \citet[][hereafter Paper I]{Raskutti2016}, we applied the
\textit{Hyperion} code in the opposite single-scattering limit to
clouds with a wide range of masses and sizes.  These simulations
showed that star formation continues unabated past the point that the
radiation force becomes super-Eddington based on a cloud's mean global
properties (i.e., in the spherical idealization).  This is because the
Eddington ratio for a given structure is inversely proportional to its
column density, and the lognormal distribution of column densities in
a cloud implies that most of the gas mass is at column densities larger
than the volume-weighted mean.  As a consequence, the net star-forming
efficiency required for radiation forces to disperse the remaining gas
can be an order of magnitude higher than would be expected based on a
spherically-symmetric idealization.  We showed that the net
star-forming efficiency in simulated clouds can be predicted
accurately, given knowledge of the mean and variance of the cloud's
column density PDF.
{The net star formation efficiency (SFE) found in Paper I
  can be considered an upper limit, since the only process that prevented
  star formation was feedback by non-ionizing FUV.  In reality, other
  internal feedback processes, as well as external dynamical effects, may
  disperse the material in a star-forming cloud
  before this ``FUV feedback-limited-efficiency'' is reached.}

Here, we consider the simulations presented in Paper I in further
detail.  We seek to understand how radiation feedback alters the
structure of stars and gas initially established by gravity and
turbulence, and how this distribution then sets the cloud disruption
times, the overall momentum input to the gas, and the statistics of
velocities in the outflowing gas.
{As in Paper I, we emphasize
  that our simulations, like other current numerical simulations, are
  not comprehensive models of star-forming clouds.  Rather, they
  should be considered controlled numerical experiments that are
  designed to focus on a particular process (here, the dynamical
  interaction between FUV radiation and cold, turbulent,
  self-gravitating gas), and to systematically analyze it under a
  range of parameterized conditions (here, principally cloud mass and
  size).

  Other forms of feedback (ionizing radiation, stellar winds,
  and supernovae from high-mass stars; prestellar outflows from
  low-mass stars) that we do not include would contribute to stirring
  up turbulence and driving outflows that could unbind cluster-forming
  clumps and clouds, and some of these may be more dynamically
  important than FUV radiation forces.  In addition, other
  idealizations that we adopt (such as not including thermal feedback,
  and omitting magnetic fields) may quantitatively affect the results,
  even if FUV radiation forces were dominant over other dynamical
  feedback.  Furthermore, realistic GMCs are never isolated, but are
  subject to larger-scale galactic processes (including nearby
  supernovae) that may act to disperse a cloud on a shorter timescale
  than internal feedback is able to be effective.

 Nevertheless,
  focused simulations and analyses of the kind we present here provide
  essential physical insight, and represent an important step towards
  future simulations that are more comprehensive and therefore better
  models of real astronomical systems.  In the present case, because
  the models of Paper I and here are the first direct numerical
  RHD simulations investigating the effects of FUV radiation pressure
  feedback in turbulent clouds, we believe it is valuable to analyze this
  process in detail before adding other physics.}
  
We begin in Section~\ref{Sec:Setup} by describing the
\textit{Hyperion} code and the numerical setup of our turbulent
clouds. In Section \ref{Sec:Fiducial} we present an overview of the
time evolution for a number of models. In
Section~\ref{Sec:Structure} we analyze the overall profiles of
gas, stars, radiation, and the Eddington force ratio in our simulations.
Section~\ref{Sec:Momentum} quantifies the statistical distributions
of gas surface density, which govern the interaction between matter
and radiation; we also measure the absorption fraction of radiation and
the total momentum imparted to the gas at different stages of evolution. 
Section \ref{sec:outflows} analyzes the 
velocity statistics of material streaming away from the central
star clusters and provides a simple interpretation based on
radiative acceleration of a lognormal surface density distribution.  
Finally, we summarize and discuss our
conclusions in the context of other theoretical work and observations
in Section \ref{Sec:Conclusion}.

\vfil

\section{Numerical Setup}
\label{Sec:Setup}

\subsection{Equations and Algorithms}

We run three-dimensional radiation hydrodynamic (RHD) simulations on a
Cartesian grid using the \textit{Hyperion} \citep{SkinnerOstriker2013}
extension to the \textit{Athena} code \citep{Stone2008}.  We refer to
\citet{SkinnerOstriker2013} for a more detailed description of the
\textit{Hyperion} code and to Paper I for details concerning the
implementation of our simulations.  However, we give here a brief
overview.

As we are operating in the single-scattering limit, we can ignore
radiative thermal emission terms, and the simplified mixed-frame equations of
RHD become:
\begin{eqnarray}
  \partial_t \rho + \nabla \cdot (\rho \mathbf{v}) &=& 0, 
\label{Eq:density} \\
  \partial_t (\rho \mathbf{v}) + \nabla \cdot (\rho \mathbf{v} \mathbf{v} + P\mathbb{I}) &=& -\rho\nabla\Phi + \rho\kappa\frac{\mathbf{F}}{c}, 
\label{Eq:RHDMomentum} \\
  \frac{1}{\hat{c}} \,\partial_t \mathcal{E} + \nabla \cdot \left(\frac{\mathbf{F}}{c}\right) &=& -\rho\kappa\mathcal{E} +\mathbb{S}, 
\label{Eq:RHDEnergy} \\
  \frac{1}{\hat{c}} \,\partial_t \left(\frac{\mathbf{F}}{c}\right) + \nabla \cdot \mathbb{P} &=& -\rho\kappa\frac{\mathbf{F}}{c}, 
\label{Eq:RHDRadMomentum}
\end{eqnarray}
where $\rho$, $\mathbf{v}$, and $P$ are the gas density, velocity, and
pressure, and $\Phi$ is the gravitational potential, all evaluated in
the lab frame. The variables $\mathcal{E}$, $\mathbf{F}$, and
$\mathbb{P}$ are the radiation energy density, flux vector, and
pressure tensor, respectively, again evaluated in the lab frame, while
$\kappa$ is the frequency-weighted specific material opacity
calculated in the gas rest frame. This opacity is set to $\kappa =
1000~{\rm cm^{2}~g^{-1}}$, consistent with the radiation pressure
cross sections per H derived from the \cite{WeingartnerDraine2001}
dust model \citep{Draine2011}.
\revis{We note that Equations (\ref{Eq:RHDEnergy}) and
  (\ref{Eq:RHDRadMomentum}) are respectively obtained as the zeroeth and
  first moments of the equation of radiative transfer, for monochromatic
  radiation.}

Our simulations adopt the simplifying assumption of an isothermal
equation of state for the gas, with $P = c_s^2 \rho$ and $c_s =
0.2~{\rm km~s^{-1}}$, corresponding to a gas temperature of $T \sim
10$~K.
{
Temperatures similar to this (or slightly higher, up to 50K)
is characteristic of most of the mass in observed GMCs
\citep[e.g.][]{ScovilleSolomon1975,Scoville1987,Roman-Duval2010,HeyerDame2015},
and is consistent with the balance between typical galactic heating rates
(by either cosmic rays or the photoelectric effect) and line cooling in
molecular gas.}
{In adopting an isothermal assumption at a temperature consistent with
  cosmic-ray heating, we ignore 
  heating by both ionizing and
  non-ionizing radiation, which would affect the gas exposed to
  unattenuated radiation from the young stars embedded in the clouds.}
As dust shielding limits the effects of
non-ionizing radiative heating to either low density material or
the cores surrounding young stars,
we assume this will have negligible impact on most of
gas in both the clouds and outflows.  Ionizing radiation may create
regions of both high temperature and pressure -- with significant
dynamical consequences -- in some parameter regimes; we shall address
the relative roles of ionizing and non-ionizing radiation in a future
publication.

\textit{Hyperion} closes the two radiation moment equations above by
adopting the $M_1$ relation \citep{Levermore1984}. This expresses the
pressure tensor in terms of $\mathcal{E}$ and $\mathbf{F}$, with
$\mathbb{P} \rightarrow (1/3)\mathcal{E}\mathbb{I}$ in the diffusion
limit ($|\mathbf{F} | / \mathcal{E}c \ll 1$) and $\mathbb{P}
\rightarrow \mathcal{E}\mathbf{\hat{n}\hat{n}}$ in the streaming limit
($| \mathbf{F} | / \mathcal{E}c \rightarrow 1$), where
$\mathbf{\hat{n}} = \mathbf{F} / |\mathbf{F}|$. As the radiation
equations are advanced in time via an explicit update,
\textit{Hyperion} ensures that the timesteps for radiation field
updates are not unfeasibly short by adopting the Reduced Speed of
Light Approximation (RSLA) \citep{GnedinAbel2001}.  In practice, for
radiation in the single-scattering limit, the
reduced propagation speed $\hat{c}$ must satisfy
$\hat{c} \sim 10v_{\rm max} \gg v_{\rm max} \sim 25-50~{\rm km~s^{-1}}$ depending 
on the initial cloud surface density 
\citep[see][and Paper I for a detailed discussion of
requirements for the RSLA]{SkinnerOstriker2013}.

Stellar emission is represented by the term $\mathbb{S}$ in
Equation~(\ref{Eq:RHDEnergy}). Monochromatic radiation from the star particles is
emitted isotropically, representing idealized luminous stellar
clusters. The source function $\mathbb{S}=j_{\rm *}/c$ for each
particle of mass $M_{\rm *}$ is Gaussian, with
\begin{equation}
	j_{\rm *}(r) = \frac{L_{\rm *}}{(2\pi \sigma_*^2)^{3/2}} \exp \left( -\frac{r^2}{2\sigma_{\rm *}^2} \right),  
	\label{Eq:jprofile}
\end{equation}
and with (fixed) radius $r_{\rm *} = \sqrt{2 {\rm ln}2} \sigma_{\rm *} =
1$~pc and fixed luminosity per unit mass $\Psi \equiv L_*/M_*$ typical
of young, luminous clusters. We adopt a fiducial value of $\Psi =
2000~{\rm erg~s^{-1}~g^{-1}}$ characteristic of a fully-sampled Kroupa
IMF \citep{Dopita2006} (again, refer to Paper I for a more
detailed discussion).
{We note that the assumption of a constant value for $\Psi$ is an
  idealization adopted for simplicity in this first study. In a
  real cluster, the light-to-mass ratio begins
  to decline significantly after $\sim 5$~Myr, whereas many of our simulations 
  have not reached completion (in the sense of either forming stars or
  driving out most of the initial gas in the cloud) within $5$ Myr from
  the onset of star formation.  
  }

{A source size $r_* = 1$pc is chosen for both physical reasons
  (this is comparable to radii of observed clusters; e.g.
  \citealt{Pfalzner2016}) and numerical reasons (this allows us to
  resolve source regions with our values of $\Delta x$).
  We note, however, that with
  our radiation solver, the radiation force will always be smaller
  than that of a point source at some scale, here $r \lsim 1$ pc.
  This implies that ongoing accretion onto star particles may be
  overestimated, even considering just the non-ionizing radiation
  (ionizing radiation would also create high-pressure non-accreting
  \ion{H}{2} regions near stars, which the present simulations do not
  capture). Nevertheless, the lack of resolution in the immediate
  neighborhood of sink particles does not affect the interaction
  between radiation and gas at larger scales, including driving
  outflows from star-forming regions, the main focus of the present study.  }

\revis{We note that the M1 closure does not allow beams of radiation to cross
(instead they merge), and more generally can fail to provide an accurate
solution for the radiation field when sources are spatially widely distributed
in an optically thin region.  For the current study, we are interested
in the interaction of the radiation field with the gas on the overall cloud
scale, whereas the individual star particles tend to be spatially concentrated.
With this source and ``screen'' geometry, 
the limitations of the M1 approximation are less
likely to affect the solution.  To check this, we have compared the solution
returned by the M1 solver with the solution computed by an adaptive ray
tracing solver (for identical source and density distribution), and
found good agreement except in the immediate vicinity of sources
(J.-G. Kim et al 2017, submitted).}

The star particles themselves are represented within the code by
point-mass sink particles \citep{GongOstriker2013} formed when cells
exceed the Larston-Penston \citep{Larson1969, Penston1969} density
threshold $\rho_{\rm th} = 8.86~c_s^2 / (\pi G \Delta x^2)$ and are
also local minima of the gravitational potential \citep{Banerjee2009,
  Federrath2010, Vazquez-Semadeni2011}. Star particles are initialized
with their mass and momentum equal to that of the gas inside a control
volume of width $3\Delta x$.  They are then evolved forward in time
using a leapfrog kick-drift-kick method \citep{Springel2005}, where
the particles' positions and momenta are alternately updated using
the particles' velocities and the total gravitational potential, 
respectively. The gravitational potential combines that of the stars
and gas and is computed via a Fourier method on a domain equivalent to
eight times the computational volume in order to implement vacuum
boundary conditions for $\Phi$ \citep{HockneyEastwood1981}.  The
density distribution input to the Poisson solver uses a particle-mesh
method with a Triangular Shaped Cloud (TSC) kernel to smoothly add each
star particle's mass to the gas density grid \citep{HockneyEastwood1981}.  
Finally, gas is accreted onto the star
particles using the HLL flux at the interface between sink
control volumes and the rest of the grid, and  sink particles are
merged when their control volumes overlap.

\textit{Hyperion} solves for the gas and radiation variables by
splitting Equations~(\ref{Eq:density})-(\ref{Eq:RHDRadMomentum}) into
separate subsystems to account for the very different time scales
involved. The gas subsystem is solved using \textit{Athena}'s unsplit
Van Leer (VL) integrator \citep{StoneGardiner2009}. The hydrodynamic
timestep is determined using a radiation-modified CFL condition with a
Courant number of $0.4$ (the typical value adopted in VL integration
schemes) and a radiation-modified effective sound speed, which accounts
for the added effect of radiation pressure in optically thick zones,
$c_{\rm eff} \equiv \sqrt{(\gamma P + 4/9\mathcal{E}(1 - {\rm
    e}^{-\rho \kappa_0 \Delta x})) / \rho}$ \citep{Krumholz2007}.

In general, \textit{Hyperion} solves the radiation subsystem using a further 
operator splitting, which separates the radiation source terms into
explicit and implicit terms. However, since we consider only radiative
absorption and not re-emission, source terms can all be solved
non-iteratively. In particular, the radiation source term $\mathbb{S}$
in Equation~\ref{Eq:RHDEnergy}, the flux absorption term
$-\rho\kappa{\mathbf{F}} / {c}$ in Equation~\ref{Eq:RHDRadMomentum}
and the radiation energy absorption term $-\rho\kappa\mathcal{E}$ in
Equation~\ref{Eq:RHDEnergy} are solved on the radiation timescale
using a standard backward Euler update. This radiation timestep is set by a CFL condition 
based on the radiation signal speed $\hat{c}$, so that there are roughly $10$
radiation substeps for each update to the gas subsystem.

\subsection{Initial Conditions}
\label{SubSec:Setup}

We consider the same set of self-gravitating star-forming clouds as
described in Paper I, evolved for $\sim 4$ initial freefall times, so
that all of the unaccreted gas is expelled from the simulation volume.  All
clouds discussed here are evolved at a resolution of $256^3$ (see Paper I
for description of convergence tests).  Each cloud is initialized as a
uniform density sphere, with ${\rho}_0 = 3M_{\rm cl,0} / (4 \pi
r_0^3)$, where $r_0$ is the initial cloud radius. The clouds are
centered inside cubic simulation volumes of length $L = 4r_0$ with
outflow boundary conditions in order to track both the mass and
momentum expelled at relatively large radius. The gas surrounding the
cloud is initialized at a factor of $10^{3}$ lower than the cloud
density, so that the total mass on the grid outside the cloud is $\sim
0.015~M_{\rm cl,0}$.

Both inside and outside the cloud, the grid is initialized with a
turbulent velocity field as described in \citet{Stone1998,
  SkinnerOstriker2015} and Paper I. This field is generated as a
Gaussian random field in Fourier space such that over the range $k \in
[2, 64] \times 2\pi / L$, $\delta v_k$ is chosen from
a Gaussian distribution with variance $P(k) \propto k^{-4}$,
consistent with observed GMCs \citep[e.g.,][]{Dobbs2013}. The velocity field is
then transformed back to real space and renormalized in terms of the
virial parameter $\alpha_{\rm vir,0} \equiv 2 E_K / |E_G|$, so that the
initial variance of the velocity distribution is $v_{\rm rms}^2 = 2 E_K / M_{\rm
  cl, 0} = \alpha_{\rm vir,0} E_G / M_{\rm cl, 0}$, where $E_K$
is the total initial turbulent gas
kinetic energy and $E_G = -3 G M_{\rm cl,0}^2 / (5 r_0)$ is the
cloud's initial gravitational binding energy.  Finally, the momentum
field is forced to have zero mean by subtracting off the initial net
momentum of the cloud. We note that this procedure results in roughly
a $2:1$ ratio of energy in solenoidal and compressive modes, respectively.

\begin{deluxetable}{cc}
\tablecaption{Fiducial Parameters\label{Tab:FiducialParams}}
\tabletypesize{\scriptsize}
\def\arraystretch{0.5}
\tablewidth{0pt}
\tablehead{
\\
\colhead{Parameter} & \colhead{Value}
}
\startdata
\\
$\alpha_{\rm vir,0}$ & 2.0 \\
$r_{\rm 0}$ & $15~{\rm pc}$ \\
$M_{\rm cl,0}$ & $5 \times 10^4~{M}_\odot$ \\
$\Sigma_{\rm cl,0}$ & $70.74~{M}_\odot~{\rm pc}^{-2}$ \\
$t_{\rm ff}$ & $4.29~{\rm Myr}$ \\
$v_{\rm rms}$ & $4.16~{\rm km~s^{-1}}$ \\
$v_{\rm esc}$ & $5.36~{\rm km~s^{-1}}$ \\
$c_s$ & $0.2~{\rm km}~{ \rm s}^{-1}$ \\
$\hat{c}$ & $250~{\rm km}~{ \rm s}^{-1}$ \\
$\Psi$ & $2000~{\rm erg}~{\rm s}^{-1}~{\rm g}^{-1}$ \\
$\kappa$ & $1000~{\rm cm}^{2}~{\rm g}^{-1}$
\enddata
\end{deluxetable}

In Table~\ref{Tab:FiducialParams}, we list simulation inputs for our
fiducial model, including the initial cloud mass, radius, and virial
parameter, and the parameters $c_s, \hat{c}, \Psi$, and $\kappa$.
In Table~\ref{Tab:ModelParams}, we show simulation
parameters and results for all models we shall discuss.  
Note that in our model nomenclature, $\Sigma$-M5E4-R15 designates
the model with initial mass $5\times10^4M_\odot$ and radius $15{\rm pc}$.
For reference, note that Table 2 in Paper I also provides initial density,
free-fall time, initial turbulent velocity dispersion, and initial
escape speed for each of these models.  All of the models considered 
here have initial virial parameter $\alpha_{\rm vir,0}=2$, which represents 
a marginally gravitationally bound state.  
For several of our analyses, we will focus on a subset of
three simulations extracted from these. We consider our fiducial model, Run
I, as well as low- and high-surface density models with $M = 2 \times
10^4~{\rm M_{\odot}}$, $R=15$pc (model $\Sigma$-M2E4-R15);
and $M = 2 \times 10^5~{\rm M_{\odot}}$, $R=15$pc (model $\Sigma$-M2E5-R15) - 
Run II and Run III, respectively - shown in italics in
Table~\ref{Tab:ModelParams}. 
For each of these models, we also consider
the case in which no radiative feedback is included (Runs Ia, IIa and
IIIa).

{
Table~\ref{Tab:ModelParams} shows that the initial free-fall time 
varies from $t_{\rm ff}=1.31~$Myr in our highest-density model to
$t_{\rm ff} = 15.3~$Myr in our lowest-density model.  As discussed in Paper I,
evolutionary times in our simulations tend to scale with the respective value of
$t_{\rm ff}$.  For example, the onset time of star formation $t_*$  varies
across the suite of models by
just 20\% upward and downward from $t_*/t_{\rm ff}=0.5$, whereas the physical
onset time varies from $t_* = 0.5 - 9.8$Myr.  In Paper I and here, we
therefore generally report times for a given simulation in units of
$t_{\rm ff}$ for that model.  To provide a sense of the corresponding physical
times, we also provide various timescales in Myr for the fiducial simulation.  

As in Paper I, we use the notation $t_{\rm x}$ to represent the time
when approximately $x \%$ of the final stellar mass has formed.  Here,
we also use the notation $t_{\rm of, x}$ to refer to the time when
$x\%$ of the outflowing gas mass has left the simulation volume.  }

\begin{deluxetable}{l|cccccccccccc}
  \tablecaption{Model Parameters and Results \label{Tab:ModelParams}}
\tabletypesize{\scriptsize}
\tablewidth{0pt}
\def\arraystretch{1.0}
\tablehead{
 \vspace{-0.2cm} &
\colhead{$\Sigma_{\rm cl,0}$} &
\colhead{$t_{\rm ff}$} &
\colhead{$t_{\rm 50}$} &
\colhead{$t_{\rm 90}$} &
\colhead{$\alpha$} &
\colhead{$f_{\rm unb}$} &
\colhead{$f_{\rm Edd}$} &
\colhead{$\sigma_{\rm ln \Sigma}$} &
\colhead{$x$} & 
\colhead{$f_{\rm abs}$} &
\colhead{$\frac{\langle p_{r, tot} \rangle }{p_*}$} & 
\colhead{$\frac{\langle p_r \rangle}{ M_*}$} 
 \\ 
\colhead{Model} \vspace{-0.2cm} &
 &
 &
 &
 &
 &
 &
 &
 &
 &
 & 
 &
 \\
 &
\colhead{[${M}_\odot~{\rm pc}^{-2}$]} &
\colhead{[${\rm Myr}$]}&
\colhead{[$t_{\rm ff}$]}&
\colhead{[$t_{\rm ff}$]}&
\colhead{} &
\colhead{} &
\colhead{} &
\colhead{} &
\colhead{} &
\colhead{} &
\colhead{} &
\colhead{[${\rm km~s^{-1}}$]}
}
\startdata
\\
$\Sigma$-M2E4-R25 & 10.13 & 14.6 & 0.88 & 1.51 & 0.84  & 0.27 & 64.58 & $1.55^{+1.23}_{-0.33}$ & $1.49^{+0.69}_{-0.27}$ & 0.37 & 0.84 & 38.28\\
$\Sigma$-M5E4-R35 & 12.92 & 15.3 & 1.07 & 2.05 & 1.16 & 0.59 & 10.24 & $1.64^{+0.23}_{-0.47}$ & $1.62^{+1.75}_{-0.45}$ & 0.39 & 1.44 & 31.79\\
$\Sigma$-M2E4-R20 & 15.83 & 10.5 & 1.08 & 1.73 & 1.00 & 0.22 & 127.16 & $1.60^{+0.27}_{-0.42}$ & $1.61^{+0.08}_{-0.42}$ & 0.44 & 0.78 & 20.64\\
$\Sigma$-M5E4-R25 & 25.33 & 9.24 & 1.01 & 1.71 & 0.92 & 0.25 & 9.82 & $1.44^{+0.01}_{-0.27}$ & $1.36^{+0.37}_{-0.20}$ & 0.50 & 1.12 & 25.72\\
$\Sigma$-M1E5-R35 & 25.84 & 10.8 & 1.00 & 1.53 & 1.20 & 0.19 & 9.85 & $1.51^{+0.02}_{-0.34}$ & $1.41^{+0.23}_{-0.25}$ & 0.62 & 2.06 & 35.39\\
  {\it $\Sigma$-M2E4-R15} & {\it 28.14} & {\it 6.79} & {\it 1.07}& {\it 1.65} & {\it 1.00} & {\it 0.14} & {\it 8.65} & {\it ${\it 1.38^{+0.05}_{-0.16}}$} & {\it ${\it 1.36^{+0.23}_{-0.18}}$} & {\it 0.53} & {\it 0.76} & {\it 15.02}\\
  $\Sigma$-M1E4-R10 & 31.66 & 5.23 & 1.09 & 1.64 & 1.05 & 0.11 & 39.70 & $1.55^{+0.10}_{-0.31}$ & $1.49^{+0.26}_{-0.29}$ & 0.45 & 0.42 & 12.92\\
  $\Sigma$-M5E4-R20 & 39.57 & 6.61 & 1.02 & 1.51 & 1.16 & 0.01 & 9.33 & $1.36^{+0.04}_{-0.18}$ & $1.28^{+0.28}_{-0.12}$ & 0.65 & 0.94 & 22.30\\
  $\Sigma$-M1E4-R08 & 49.46 & 3.74 & 1.11 & 1.75 & 1.02 & 0.44 & 37.57 & $1.52^{+0.15}_{-0.24}$ & $1.43^{+0.70}_{-0.22}$ & 0.45 & 0.35 & 9.25\\
  $\Sigma$-M1E5-R25 & 50.65 & 6.54 & 1.05 & 1.48 & 1.14 & 0.01 & 8.68 & $1.36^{+0.06}_{-0.19}$ & $1.27^{+0.18}_{-0.11}$ & 0.66 & 1.06 & 26.50\\ 
  $\Sigma$-M2E5-R35 & 51.69 & 7.66 & 1.08 & 1.51 & 1.28 & 0.10 & 7.28 & $1.45^{+0.02}_{-0.27}$ & $1.34^{+0.26}_{-0.18}$ & 0.70 & 1.66 & 34.43\\
  $\Sigma$-M5E3-R05 & 63.31 & 2.61 & 1.17 & 1.95 & 0.91 & 0.05 & 34.53 & $1.65^{+0.10}_{-0.33}$ & $1.59^{+0.91}_{-0.35}$ & 0.27 & 0.28 & 6.41\\
  $\Sigma$-M2E4-R10 & 63.31 & 3.70 & 1.21 & 1.83 & 0.86 & 0.02 & 6.19 & $1.72^{+0.14}_{-0.20}$ & $1.61^{+0.54}_{-0.25}$ & 0.49 & 0.40 & 10.70\\
  {\bf $\Sigma$-M5E4-R15} & {\bf 70.35} & {\bf 4.29} & {\bf 1.06} & {\bf 1.57} & {\bf 1.16} & {\bf 0.03} & {\bf 13.21} & {\bf ${\bf 1.39^{+0.05}_{-0.24}}$} & {\bf ${\bf 1.27^{+0.37}_{-0.13}}$} & {\bf 0.61} & {\bf 0.60} & {\bf 16.41}\\
 $\Sigma$-M1E5-R20 & 79.14 & 4.67 & 1.02 & 1.56 & 1.13 & 0.03 & 4.30 & $1.41^{+0.02}_{-0.29}$ & $1.27^{+0.32}_{-0.14}$ & 0.70 & 0.95 & 21.14\\  
  $\Sigma$-M2E4-R08 & 98.93 & 2.64 & 1.10 & 1.71 & 1.09 & 0.02 & 7.49 & $1.44^{+0.11}_{-0.21}$ & $1.32^{+0.39}_{-0.15}$ & 0.44 & 0.31 & 8.50\\
  $\Sigma$-M2E5-R25 & 101.30 & 4.62 & 1.04 & 1.57 & 1.17 & 0.02 & 4.70 & $1.43^{+0.05}_{-0.33}$ & $1.28^{+0.31}_{-0.17}$ & 0.76 & 1.17 & 28.09\\
  $\Sigma$-M1E4-R05 & 126.63 & 1.85 & 1.16 & 1.80 & 1.14 & 0.02 & 5.71 & $1.55^{+0.04}_{-0.26}$ & $1.42^{+0.35}_{-0.21}$ & 0.23 & 0.22 & 5.93\\
  $\Sigma$-M1E5-R15 & 140.70 & 3.04 & 1.04 & 1.63 & 1.10 & 0.03 & 4.75 & $1.36^{+0.14}_{-0.25}$ & $1.24^{+0.45}_{-0.12}$ & 0.68 & 0.50 & 16.49\\
  $\Sigma$-M5E4-R10 & 158.29 & 2.34 & 1.09 & 1.67 & 1.12 & 0.04 & 4.19 & $1.39^{+0.08}_{-0.24}$ & $1.27^{+0.33}_{-0.13}$ & 0.55 & 0.34 & 9.93\\
  $\Sigma$-M2E5-R20 & 158.29 & 3.31 & 1.09 & 1.65 & 1.12 & 0.01 & 4.19 & $1.31^{+0.07}_{-0.21}$ & $1.21^{+0.34}_{-0.10}$ & 0.77 & 0.64 & 19.96\\
  $\Sigma$-M5E4-R08 & 247.32 & 1.67 & 1.11 & 1.77 & 1.13 & 0.03 & 4.49 & $1.37^{+0.11}_{-0.23}$ & $1.27^{+0.32}_{-0.14}$ & 0.44 & 0.30 & 9.30\\
  $\Sigma$-M2E4-R05 & 253.26 & 1.31 & 1.20 & 1.86 & 1.09 & 0.01 & 4.31 & $1.48^{+0.12}_{-0.30}$ & $1.35^{+0.29}_{-0.21}$ & 0.20 & 0.18 & 6.36\\
{\it $\Sigma$-M2E5-R15} & {\it 281.40} & {\it 2.15} & {\it 1.22 } & {\it 2.01} & {\it 0.91} & {\it 0.00} & {\it 3.31} & {\it ${\it 1.37^{+0.24}_{-0.21}}$} & {\it ${\it 1.28^{+0.46}_{-0.10}}$} & {\it 0.73} & {\it 0.37} & {\it 15.73}\\   
  \enddata
\tablecomments{Columns display the following information
  (i) model name,
  (ii) initial cloud surface density,
  (iii) initial free-fall time,
  (iv) time $t_{\rm 50}$ when  $50\%$ of star formation is complete,
  (v) time $t_{\rm 90}$ when  $90\%$ of star formation is complete,
  (vi) power law exponent of the fitted density profile at $t_{\rm 50}$
  (vii) fraction of the stellar mass unbound by the simulation end,
  (viii) Eddington factor, defined in Equation~\ref{Eq:FEdd}, at $t_{\rm 90}$,
  (ix) width of the lognormal circumcluster surface density distribution
  at $t_{\rm 50}$ with error bars given at $t_{\rm 10}$ and $t_{\rm 90}$ respectively,
  (x) cloud size factor $x$, defined in Equation~\ref{Eq:x}
  at the same times as for $\sigma_{\rm ln \Sigma}$, 
  (xi) radiation absorption fraction measured 3 Myr after star formation begins,
  (xii) total radial gas momentum in the simulation volume 
  plus outflowing divided by
  input stellar radiation momentum, measured 3 Myr after star formation begins,
  (xiii) final outflowing gas momentum per unit of stellar mass formed.
  The fiducial model (Run I) is shown in bold ($\Sigma$-M5E4-R15), while the low and high 
  surface density models (Runs II and III) are shown in italic
($\Sigma$-M2E4-R15 and $\Sigma$-M2E5-R15).}
\end{deluxetable}

\section{Overview of Time Evolution -- Gas and Radiation Structure}
\label{Sec:Fiducial}


\begin{figure*}
  \centering
  \epsscale{1}
  \includegraphics{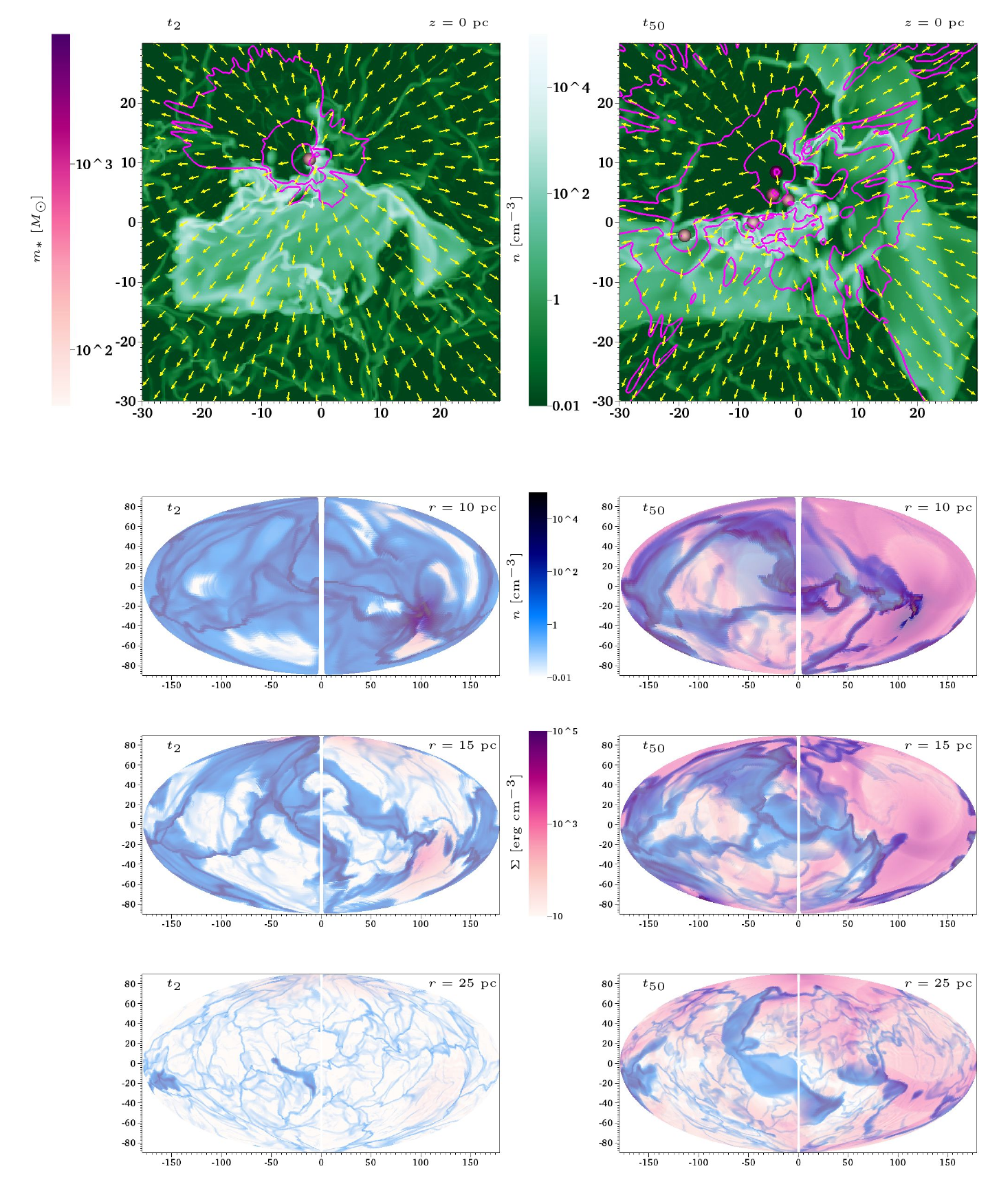}
  \caption{Snapshots of the gas density for the fiducial model at
    times $t_{\rm 2}/ t_{\rm ff} = 0.44$ (left),~ and
    $t_{\rm 50}/ t_{\rm ff} =1.04$ (right).
{
    In physical units,  these times are 
    $t_{\rm 2} = 1.9$ Myr and $t_{\rm 50} = 4.5 $ Myr, when respectively $2\%$
    and $50\%$ of the final stellar mass in the fiducial model has formed.}
We show (top) slices through
    the x-y plane passing through the position of the stellar center
    of mass, and (bottom) Hammer projections of spherical slices
    around that same center, at radii of $10$, $15$, and $25$~pc. In
    the x-y slices, density is shown with a green color scale (units
    ${\rm cm^{-3}}$, top middle), the directions of radiation flux vectors are
    overlaid in yellow, and pink contour lines, separated by decades
    relative to the peak value, show the radiation energy density.
    Star particles within $\Delta z = \pm 2$~pc of the slice are
    plotted as circles, with color scale for the particle mass (red) 
    in units of $M_{\odot}$, shown at the top left. 
    In the Hammer projections, density is shown with a blue color scale
  (in units of ${\rm cm^{-3}}$), and radiation energy density is overlaid in pink.}
  \label{Fig:FiducialMaps}
\end{figure*}

We begin by considering the time evolution of the fiducial model Run I
(as described in Section~\ref{Sec:Setup}).  
In Figure~(\ref{Fig:FiducialMaps}) we 
show density maps for Run I, viewed
both as slices through the stellar center of mass and as Hammer projections 
of spherical slices around the same center and at various radii. 
We show slices, with contours of radiation energy density overplotted, 
at times $t_{\rm 2}=0.44t_{\rm ff}=1.9$Myr (left) and
$t_{\rm 50}=1.04 t_{\rm ff}=4.5$Myr (right).

At the earlier time ($t_2/t_{\rm ff}=0.44$),
the fiducial cloud has already collapsed, with
mass gathered by turbulence and self-gravity
preferentially along two perpendicular filaments,
although reasonably dense material still extends out to several pc
surrounding these filaments. The Hammer projections show evidence of
the filamentary nature of the cloud and of an overdensity where a
nearby star is forming, but there are relatively few holes in the cloud inside
of the original cloud radius (15~pc).

However, by the time half the stars have formed ($t_{\rm 50}/t_{\rm ff} = 1.04$),
the structure looks
quite different. Radiative feedback has pushed the surrounding gas
into much thinner, denser filaments, such that
very little of the sky is covered by absorbing
gas. Thus, the majority of radiation may freely stream away from sources
without imparting momentum to a significant fraction of the gas mass. We note that the
structure seen here is quite different from the structure seen in
\citet{SkinnerOstriker2015}, where only diffuse IR radiation is included.  In that
case, the
lower opacity for IR compared to UV means that radiation forces are
more distributed within the gas (creating less intense compression of
filaments) and low-density regions are not as rapidly cleared.

Although the dominant characteristic of cloud structure is the
filamentary morphology, clouds also becomes radially stratified, as will
be discussed in Section~\ref{Sec:Structure} below.  Because there is
no re-emission or scattering of absorbed radiation, the flux is strongly
anticorrelated with the circumstellar surface density; this is also
evident in the anticorrelation of radiation energy density with gas
density.  Furthermore, since the dense filaments cover very little
solid angle, they are ineffective in shielding low density gas at
large distance from the center, so this gas is rapidly cleared by
radiation forces.

All of our simulation runs have the same initial spatial distribution
from the initial turbulent velocity field and differ only in the
amplitude $v_{\rm turb}$, so that qualitatively similar gas structures and time
evolution are evident in both the low-$\Sigma$ and high-$\Sigma$
models (Run II and Run III) compared to the fiducial model (Run I).
In the low-$\Sigma$ run, radiation is very efficient at dispersing
cloud material and only $\sim 10\%$ of the initial cloud is converted
to stars; the high-$\Sigma$ run has a much higher final SFE $\sim 60\%$.
In all cases, the tenuous material between filaments is expelled first,
and as a consequence much of the radiation escapes through large holes
in the cloud, even in the high-$\Sigma$ model.

\section{The Profiles of Gas, Stars, and Radiation in the Cloud}
\label{Sec:Structure}

As discussed, the morphology of our simulated clouds is highly
filamentary, with star formation concentrated in the filaments.
Nevertheless, a number of key insights can be gained by considering
the angle-averaged profiles of the cloud in spherical shells.
Most galaxy formation simulations still struggle to resolve GMC
scales, and in some cases \citep[e.g.,][]{Olsen2015} where chemistry
and emission properties are of interest, clouds are instead modeled
via a sub-grid prescription with an adopted spherical density profile.
Meanwhile, numerical or semi-analytic models that combine the effects
of different types of feedback \citep[e.g.,][]{Sales2014} often rely
on assumptions of spherical symmetry for star-forming clouds. As
results in these and other applications can depend strongly on the
assumed cloud density profile, knowledge of the angle-averaged
spherical structure established by turbulence and feedback is
desireable.

The radial distribution of gas and stars in our simulations is also
interesting, because the boundedness and eventual mass distribution of
observed stellar clusters depends strongly on how centrally concentrated
embedded clusters are within GMCs. Observations suggest that close to
$\sim 90\%$ of all stars formed in molecular clouds are eventually
unbound \citep{LadaLada2003}, while simulations indicate that the
bound fraction depends sensitively on the relative shape of the gas
and stellar potentials prior to gas expulsion by feedback
\citep{Lada1984, Goodwin1997, Adams2000, GeyerBurkert2001,
  BoilyKroupa2003, GoodwinBastian2006, BaumgardtKroupa2007,
  ProszkowAdams2009}.  In particular, if clusters are more centrally
concentrated with a high local SFE, the eventual feedback-driven gas expulsion
from large radii will have limited effect on unbinding the clusters
at small radii \citep[see, e.g.,][for a detailed discussion]{Longmore2014}.

\subsection{Gas Profiles}
\label{SSS:profile}

\begin{figure}
  \centering
  \epsscale{1}
  \includegraphics{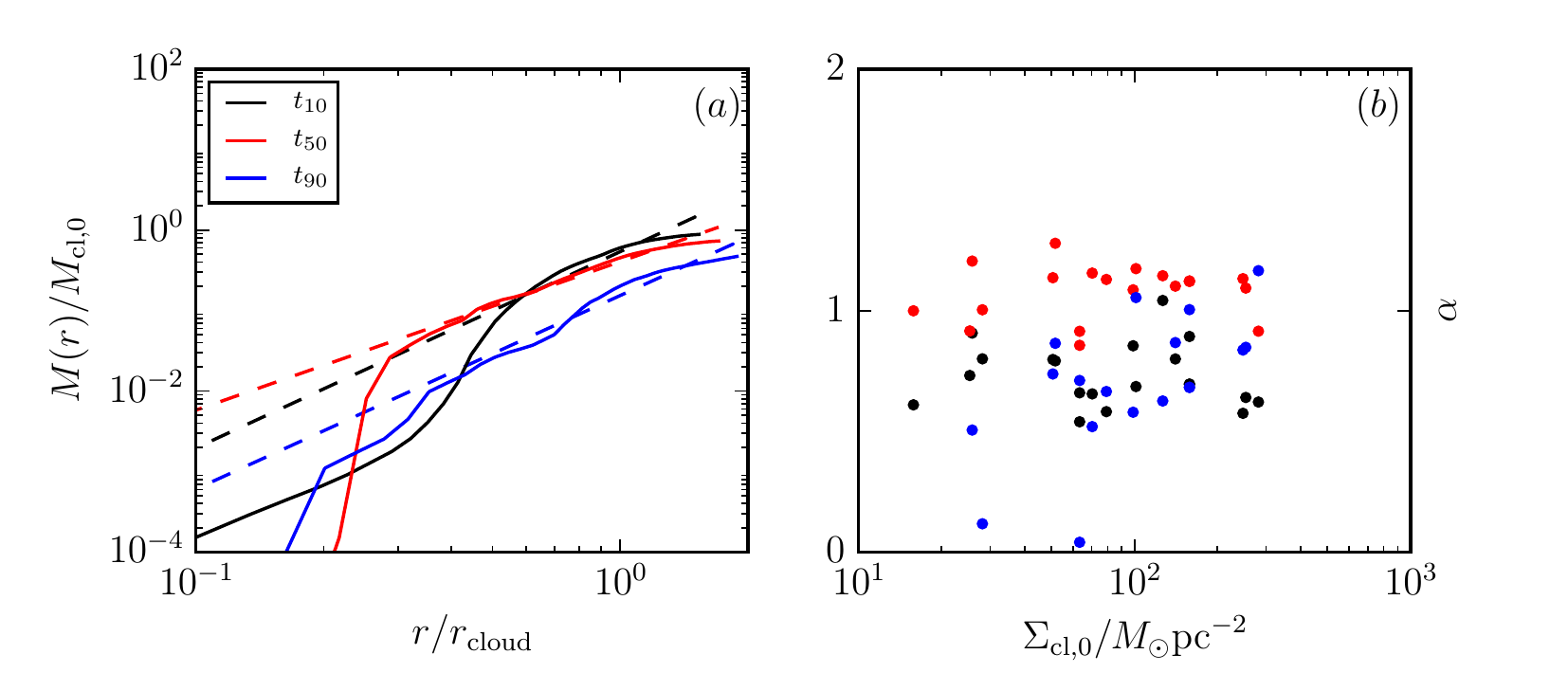}
  \caption{Angle-averaged radial profiles of the mass 
  $M(r) = \int_0^r dr 4 \pi r^2 \langle \rho \rangle$ enclosed
    within a sphere of radius $r$ (normalized to $M_{\rm cl,0}$)
    at three different times, denoted by $t_{\rm x}$ when the stellar efficiency ($\varepsilon$)
    is at $x\%$. On the left, we show profiles for the fiducial
    model with radiation feedback, where at each time, 
    we also show the best-fit power law (dashed) over
    the range between $r=3$ and $15$~pc. On the right, we show 
    best-fit values of the power law exponent $\rho(r) \propto r^{-\alpha}$ as a function of 
    initial cloud surface density for all models in the $\Sigma$-series.}
  \label{Fig:RhoProfile}
\end{figure}

\begin{figure}
  \centering
  \epsscale{1}
  \includegraphics{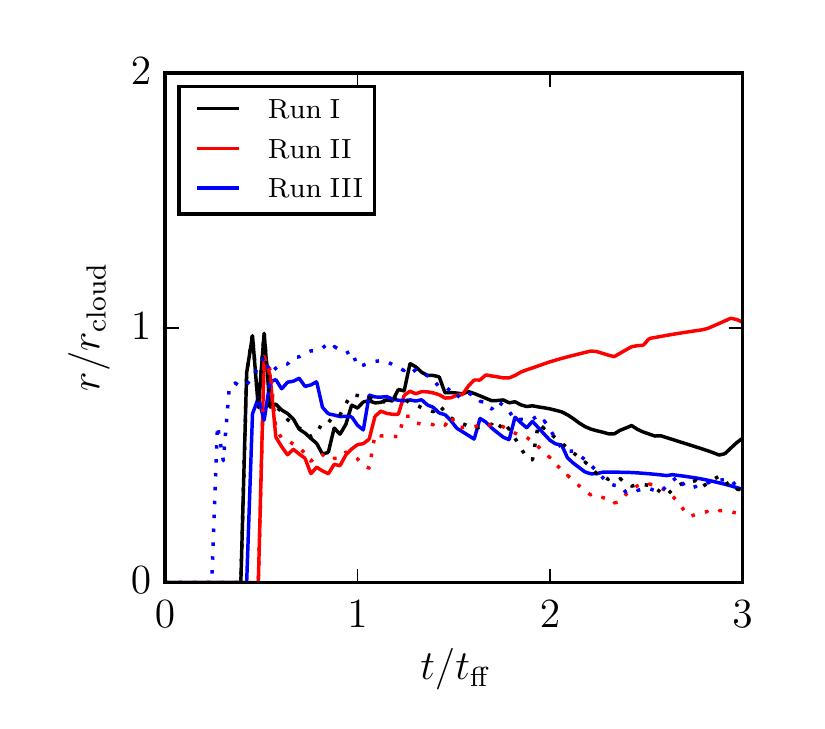}
  \caption{Mean size of the stellar mass distributions as a function of time 
  in Runs I-III (solid lines) as well as Runs Ia-IIIa (without feedback, dotted lines). In each case, 
  we show the radius which contains $68~\%$ of the mass.}
  \label{Fig:RadProfilevsT}
\end{figure}

{In describing radial profiles of the cloud, we choose as our coordinate 
  center the center of mass of the system of star particles.  We make
  this choice because we are primarily interested in the interaction of
  radiation with the gas, and with an (adopted) constant light-to-mass
  ratio the center of the luminosity source is this center of mass.}
  After identifying this center, we interpolate from 
the Cartesian grid onto a grid in $[r, \theta, \phi]$
space, where the grid points are spaced equally in $r$, $\phi$, and
${\rm cos} \theta$.
We then compute angle averaged radial profiles, denoted by
$\langle . \rangle$, averaging over all $\theta$ and $\phi$ at a given
radius.  In Figure~\ref{Fig:RhoProfile}a we show radial profiles of gas
mass enclosed in spherical shells at several different times
throughout the fiducial cloud's evolution.  As before, times are denoted 
by $t_{\rm x}$ when the stellar efficiency ($\varepsilon$)
is at $x\%$.
For the fiducial model, by the time $10\%$ of
{the final stellar mass has formed
($t_{\rm 10}=0.56 t_{\rm ff}=2.4$ Myr)} we find that the cloud's mass
profile follows a roughly power-law shape, implying that the
underlying gas density profile has itself settled into a rough power
law.  The shape of the profile does not change significantly until the
very end of star formation, when direct radiation pressure drives strong
outflows, compressing the remaining gas into a thin shell.

Parameterizing the density profile as a time-varying power law,
$\rho(r) \propto r^{-\alpha}$, we find that $\alpha \sim 1.2$ at $t_{\rm 50}$
for the fiducial model.
Figure~\ref{Fig:RhoProfile}b shows that similar results hold true
over the full range of clouds that we model, with $\alpha \sim 1$
at $t_{50}$.  In Table~\ref{Tab:ModelParams}, the values of $t_{\rm 50}$
and $\alpha$ at $t_{\rm 50}$ are listed for all models.  
We note that $\alpha \sim 1$ 
is shallower than the isothermal or Plummer
spheres \citep{Plummer1911, WhitworthWard-Thompson2001} often 
assumed for GMCs \citep{Sales2014, Olsen2015}
or seen in observations of the central star-forming core of the Orion
Nebula Cluster \citep{DaRio2014}.
It is also shallower than the $\alpha=1.5$
exponent found in recent simulations \citep{Lee2014}, which do not include feedback.
However, we also 
note that we fit the power law over a large range of radii, and that we
lack the resolution necessary to fit the central star-forming
region well (in fact, the density profiles do appear steeper in
the innermost region).

If we follow the argument of \cite{MckeeTan2003} that the star
formation rate scales with the inverse of the freefall time ($\sim \rho^{-1/2}$)
of freely collapsing cores, this would lead to an SFR $\propto t^{6/\alpha -3}$,
which would suggest a steep increase in time for
$\alpha \sim 0.8-1.3$.
Instead, as discussed in Paper I, we
observe a roughly linear SFR across all surface densities.
We believe that this is because the locations (and rate)
of collapse are primarily
controlled by the filamentary gas structure, and only secondarily by
overall cloud stratification.  

\subsection{Stellar Distribution Profiles}

We begin this section by comparing the relative sizes of the nascent stellar clusters
and the clouds in which they are embedded. The spatial distribution of star
particles has a long tail, since a small number become
unbound and escape the simulation volume. We
therefore define the size of the cluster as the $1$-$\sigma$ mass range,
i.e. the radius of the sphere about the stellar center of mass containing $\sim 68~\%$ of 
the stellar mass.
Figure~\ref{Fig:RadProfilevsT} compares the evolution in the radius of
the stars and gas for Runs I, II, and III, as well as the corresponding Runs Ia, 
IIa, and IIIa without radiation.

The stellar distributions evolve in very different ways for the
three runs. At early times,
star formation is far more centrally
concentrated in the low-surface-density Run II,
compared to the somewhat larger 
radial extent in the case of Run III. However, the extended accretion
process in Run III  binds the cluster more tightly, and the stellar
distribution eventually shrinks after $t\sim t_{\rm ff}$. By $t \sim 1.5 t_{\rm ff}$,
the effective stellar radii of all runs are roughly the same.
At late time, the rapid
gas expulsion in Run II appears to unbind the cluster as well, its radius
increasing secularly in time.  For the fiducial model (Run I), the effective
stellar radius shrinks after $t \sim 1.5 t_{\rm ff}$.

Figure~\ref{Fig:RadProfilevsT} also shows the evolution of stellar distribution
in Runs Ia-IIIa, the no-feedback comparison simulations.  The most significant
difference is that the cluster distribution remains bound in Run IIa,
the low-surface density no-feedback model.

{
Other low-$\Sigma$ runs show behavior similar to the Run II, with the
stellar distribution expanding over time, and a relatively large fraction
of the stellar mass unbound at the end of the simulation (see $f_{\rm unb}$ in
Table~\ref{Tab:ModelParams}).  Simulations with higher $\Sigma$ have a
very small fraction of the star particles unbound at late times.  This 
suggests that the evolution of the star particle distribution in
our simulations is primarily a response to the net SFE 
in the clouds (which is controlled by radiation pressure feedback),
and is low in low-$\Sigma$ clouds and high in high-$\Sigma$ clouds.  

Finally, we caution that due to limited resolution, our star particles
do not represent individudal stars, so the late-time dynamics
may be quite different from that of a real cluster of the same mass.
}

\subsection{Radiation Profiles}\label{Sec:radprof}

\begin{figure}
  \centering
  \epsscale{1}
  \includegraphics{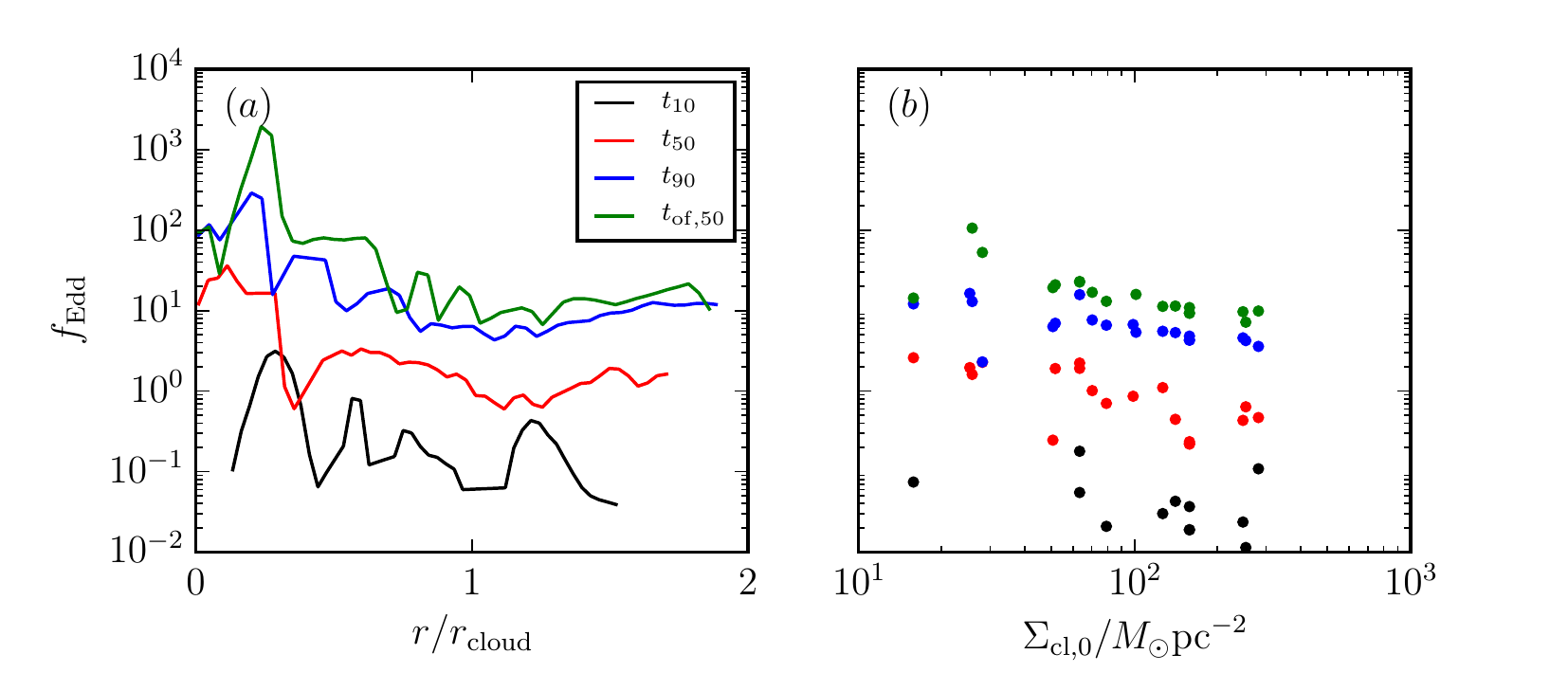}
  \caption{Angle averaged Eddington factor, as
    described in Equation~(\ref{Eq:FEdd}). We show profiles for the fiducial
    model (left) at four different times when the
    stellar efficiency or outflow fraction is as shown in the key
    (see text).  For the same
    respective times in each model, on the right 
    we show the Eddington factor at the cloud radius as a function
    of initial cloud surface density. 
    We note that, particularly at small
    radii, the Eddington factor may be negative as star particles just
    exterior to a given shell may dominate the flux.
    For simplicity, we therefore omit these radii in $(a)$. }
  \label{Fig:FEdd}
\end{figure}

We now turn to
an exploration of
the clouds' internal radiation profiles, focusing
on the competition between radiation and gravitational forces. We do
this, as before, using angle-averaged spherical profiles 
interpolated from the Cartesian grid.
{
The origin of the spherical
coordinate grid is at the center of luminosity of the star particle
distribution, so that outside of both the stellar and gas distribution
(beyond $\sim 0.8 ~r_{\rm cloud}$) the angle-averaged flux decreases
$\propto r^{-2}$.}

We may directly compare the influences at different radii of the centrally 
concentrated radiation sources, and the more distributed gravitational
potential, 
by considering the Eddington factor, $f_{\rm Edd}(r)$. 
The mean Eddington factor in a given spherical shell is defined as 
the ratio of the radiation to gravitational forces, given by 
\begin{equation}
	f_{\rm Edd}(r) = \frac{\langle \rho \kappa F_r / c
          \rangle}{\langle \rho \partial_r \Phi \rangle}.
	\label{Eq:FEdd}
\end{equation}
For isotropic distributions of gas density and radiation, the
force on each fluid element within a given radial shell would be
identical.  In this case, neglecting hydrodynamic stresses, 
we would expect star formation to
continue until $f_{\rm Edd}(r)=\kappa L(r)/[4 \pi G M(r)]$ exceeds unity
everywhere in the cloud; here $L(r)$ and $M(r)$ are the total luminosity
and mass within $r$.

Hydrodynamic stresses lead to internal momentum exchange but 
average to zero over a sufficiently large volume.  Thus, it is also interesting 
to consider the cumulative Eddington factor, defined by 
\begin{equation}
	f_{\rm Edd,cum}(r') = \frac{\int_0^{r'} r^2 \langle \rho \kappa F_r / c
          \rangle dr}{\int_0^{r'} r^2 \langle \rho \partial_r \Phi \rangle dr},
	\label{Eq:FEddCum}
\end{equation}
which represents the ratio of the volume-averaged radiation 
to gravitational forces out to a given radius.  
In the isotropic case, we might expect star 
formation to slow  when $f_{\rm Edd,cum}(r_{\rm cl}) \sim 1$.
In fact, as discussed in Paper I, the distributions of gas and radiation in the cloud are far from
isotropic (or uniform), so that some fluid elements within a given shell may
be sub-Eddington even when the angle-averaged Eddington factor exceeds
unity.  As a consequence, star formation can continue locally even when the mean 
or cumulative Eddington factor at the cloud radius is well above unity.  

Figure~\ref{Fig:FEdd}$a$ shows the radial profiles of the
angle-averaged Eddington factor $f_{\rm Edd}(r)$ for our fiducial
model.
{ We show profiles from $t_{\rm 10}/t_{\rm ff}=0.56$,
  $t_{\rm 50}/t_{\rm ff}=1.06$, $t_{\rm 90}/t_{\rm ff}=1.57$, and
  $t_{\rm of, 50}/t_{\rm ff}=2.17$, where $t_{\rm of,50}=9.3$ Myr
  represents the time when 50\% of the mass has been driven out of the
  cloud by radiation forces.\footnote{
{    We note the important caveat
    that during the later
    evolutionary stages of the simulation,
    the timescales are long enough that in a real
    cloud, the radiation field would have declined and the most
    massive stars would have exploded as supernovae. The present
    idealized simulations do not include these effects.}
}
  }

Since
$r^2 \langle F_r\rangle $
flattens at large radii (because stars are centrally concentrated, and
luminosity is attenuated) while
$r^2\langle \partial_r \Phi\rangle$ continues to increase (from
the distributed gas), $f_{\rm Edd}(r)$ is expected to be lowest at large
radii.  Except at the very beginning of star formation, Figure~\ref{Fig:FEdd} indeed shows
that the Eddington factor is largest at small radii and declines outward.
Beyond $\sim 0.5~r_{\rm cl,0}$, we find that the local and 
cumulative Eddington factors are approximately equal, particularly at later times, 
approaching a nearly constant value beyond $r_{\rm cl,0}$.
Figure~\ref{Fig:FEdd}$b$ shows that all of our models reach
$f_{\rm Edd}(r_{\rm cl,0}) \sim 1$ at $\sim t_{\rm 50}$
when half of their stars are
formed, but continue forming stars well beyond this point.
By the time star formation is nearly over, $f_{\rm Edd}(r_{\rm cl,0})$
is an order of magnitude higher.

We note (see Figure~\ref{Fig:FEdd}$b$)
that there is a slight downward trend with initial cloud
surface density in the mean 
Eddington factor at all stages of star formation. In particular, the 
limiting (i.e., at 90\% complete star formation or
50\% complete outflow ejection) mean Eddington factor
$f_{\rm Edd}(r_{\rm cl,0})$ appears 
to decrease by a factor 5 or more from $\Sigma \sim 10~M_{\odot}~{\rm pc^{-2}}$ to 
$\Sigma \sim 200~M_{\odot}~{\rm pc^{-2}}$.
Therefore, there  
does not seem to be a single, global Eddington factor at which clouds 
are destroyed. However, we note that in all of our simulations,
$f_{\rm Edd}(r_{\rm cl,0})$ at this
late stage is at least an order of magnitude larger than unity.
{
  As we showed in Paper I (see also below), the truncation of star formation
  by radiation pressure in a turbulent cloud depends on the variance of
  the surface density distribution, so the range of final $f_{\rm Edd}(r_{\rm cl})$
  that we find depends on this variance. The final value of
  $f_{\rm Edd}(r_{\rm cl})$ would be lower if the 
  surface density variance is lower than it is in our models, which may
  be true in real clouds. 
  }

\section{Statistics of Cloud Structure and Interaction with Radiation}
\label{Sec:Momentum}

In Section \ref{Sec:radprof}, we showed that when the angle-averaged
Eddington factor reaches a typical value $\sim 10$ at the cloud radius, star formation
shuts down and the remaining gas is ejected from the cloud.  This
order of magnitude increase in $f_{\rm Edd}(r_{\rm cl,0})$ 
(relative to the simple spherically-symmetric prediction)
mirrors the large increase in the net
SFE (compared to the spherical prediction)
we found in Paper I.  In that work, we argued that the
boost in SFE may be attributed to the lognormal distribution of
surface densities in the circumcluster gas: 
high surface density regions remain bound even when the
average Eddington factor exceeds unity.  Thus, to quench star formation, 
the luminosity must increase (by forming additional
stars) to such a level that even the high-$\Sigma$ tail of the lognormal distribution
becomes super-Eddington.

In Paper I, we also showed that the surface density mass PDF
(i.e. the distribution of mass as a function
of surface density $\Sigma$ or column densities $N=\Sigma/\mu$),
$P_M(\Sigma)$,
generally has a lognormal shape over the main period of star formation
$t/t_{\rm ff} \sim 0.5 -1.5$. Similarly, the 
surface density area PDF
(i.e. the distribution of area as a function
of surface density $\Sigma$),
$P_A(\Sigma)$, also follows a lognormal at the high-$\Sigma$
end, but because the computational domain  contains a large, low-density
volume outside the cloud, there is a low-$\Sigma$ excess above the lognormal 
from the ``non-cloud'' material.  We showed that 
$\sigma_{\ln \Sigma}$ of $P_M$ and the ``cloud'' portion of $P_A$ are the same,
as expected.  Also as expected, the means are given by 
$\langle \ln(\Sigma/\overline\Sigma_{\rm cloud})\rangle_{M,A} = \pm\sigma_{\ln \Sigma}^2/2$.
In that work, we also showed that while the mean surface density of
clouds changes significantly over time as gas is consumed by star formation and ejected
by radiation pressure, the value of $\sigma_{\ln \Sigma}$ evolves much less,
and is similar for all of our models.  

While in Paper I we analyzed the distribution of 
surface densities as
would be measured
{in a Cartesian projection} by an external observer, the distribution from the point of view of 
the embedded stars is more relevant when
considering the interaction of the gas with radiation.  As shown in that work, denoting 
the current SFE as $\varepsilon$, structures with circumcluster
surface density $\Sigma^c$ below
$\Sigma_E=\varepsilon(1+\varepsilon)^{-1}\Psi(2\pi c G)^{-1}$ will be
super-Eddington, hence can be driven out of the cloud, while structures
with $\Sigma^c > \Sigma_E$ cannot.  In Paper I, we developed a
formalism that provides a prediction for the final SFE in a cloud
based on the value for which the largest possible fraction of gas has
$\Sigma^c < \Sigma_E$ and is ejected. We showed that this predicted
SFE is in good agreement with the measured final SFEs in our
simulations. To provide further quantitative support for the formalism
and SFE predictions of Paper I, here we analyze the {\it circumcluster}
surface density distributions for our cloud models.

In addition to determining the fraction of a cloud's gas mass that is
super-Eddington, the circumcluster surface density distributions
are also important for determining the fraction of radiation that
escapes from the cloud. Finally, since the acceleration of a structure
by radiation pressure depends on its surface density, the total momentum and
distribution of outflowing gas velocities depend on the distribution
of circumcluster surface densities.  We shall start in
Section \ref{SSS:Sigma_evol} by comparing the
evolution of observed and circumcluster surface densities,
and then in Sections \ref{SSS:PDF} and \ref{SSS:escape}
provide quantitative measures for their lognormal distributions and
for the escape of radiation in all models.  In Section \ref{sec:outflows} we
shall connect the properties of outflowing gas to the statistics
of these circumcluster surface densities.

\subsection{Evolution of Observed and Circumcluster Surface Density}
\label{SSS:Sigma_evol}

\begin{figure*}
  \centering
  \epsscale{1}
  \includegraphics{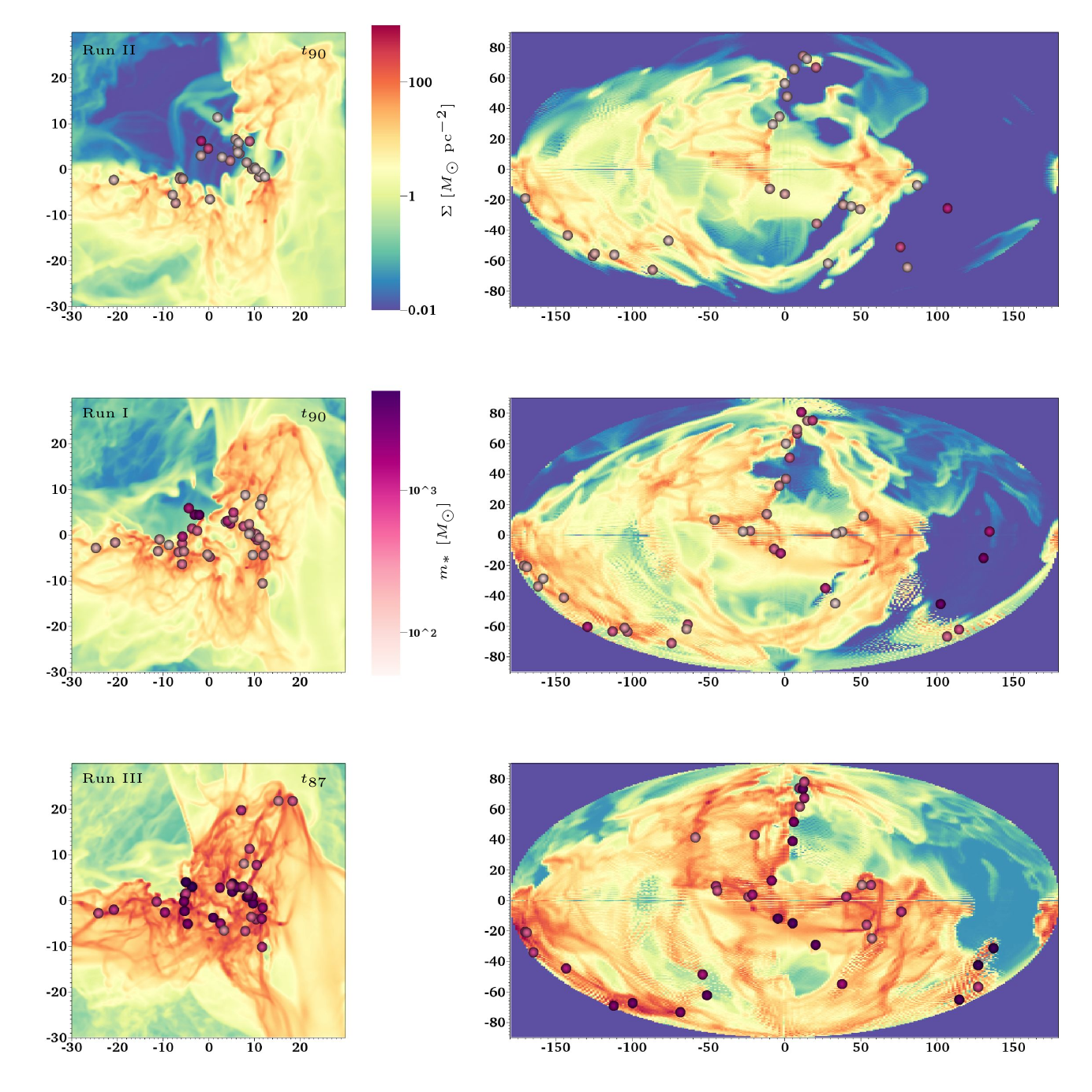}
  \caption{Snapshots of the surface density at time
    $t = 1.57~t_{\rm ff} = t_{\rm 90}$ for Run I (middle). We also show snapshots
    for the low-$\Sigma$ Run II (top) 
    and the high-$\Sigma$ Run III (bottom) at $t=1.57 t_{\rm ff}$ for their
    respective free-fall times, when the 
    stellar mass is at $90 \%$ and $87 \%$, respectively. We show both projections along the z-axis onto the
    x-y plane (left), as well as Hammer projections of the column
    density in rays starting $1.2~$pc away from the stellar center of
    mass and continuing until the sphere intersects the outer edge of the box (right). 
    The inner radius is chosen to omit a region of $3^3$ voxels around the center of mass, 
    while the outer radius is chosen such that all rays have the same radial extent, 
    although some do not reach the edge of the simulation volume. 
    The color scales show surface density in units of $M_{\odot}~{\rm pc^{-2}}$ (top) 
    and star particles as spheres, 
    with the color scale (middle) showing the stellar mass in units of
    $M_{\odot}$.}
  \label{Fig:LateSDMaps}
\end{figure*}

At any timestep, we can obtain the circumcluster surface density distribution 
by following the procedure
outlined in Section~\ref{Sec:Structure}.  For each point on a grid
in $[r, \theta, \phi]$, we draw rays from the stellar center of mass
to the edge of the box. We then calculate the integral of
density along each ray as $\Sigma^c = \int \rho dr$.
We note that although $\Sigma^c$ has units of mass/area, it 
does not correspond exactly to a surface density except in the
special case of a thin shell, since the area of each cell in the
$[r, \theta, \phi]$ grid 
increases radially outward as $r^2$. However, along any given line of
sight, if the gas is concentrated over a relatively small radius
range, $\Sigma^c$ represents a local surface density as seen by a central
source.  For a constant opacity $\kappa$, the optical depth along
a given ray is $\Sigma^c \kappa$.

In the middle row of
Figure~\ref{Fig:LateSDMaps} we show both the ``observed'' surface density map
(projected along $z$) and the Hammer projection of the
circumcluster surface density in the fiducial model (Run I) at $t = 1.57~t_{\rm ff}$, 
when $90 \%$ of the final stellar mass has
formed.  Qualitatively, the maps are quite similar in that gas is
concentrated in large, dense filaments, which are also the sites of star 
formation.  The top and bottom rows of Figure~\ref{Fig:LateSDMaps}
shows corresponding maps for Runs II and III.

However, at this late time, the differential effects of radiative feedback are
evident. The most massive star (the farthest east along the equator)
and its neighbors have by this stage created a blowout, clearing gas
out of the top-left corner of the ``observed'' clouds and a large hole
in the east of the Hammer projections.  The blowout (and other
smaller, local bubbles) becomes more pronounced with decreasing
surface density (Run I). This reflects the different stages of these
clouds' evolution; the high-density model (Run III) continues to
accrete gas onto stars until radiative feedback from the most massive
star becomes strong enough to evacuate its local environment.

\subsection{Gas Density PDF}
\label{SSS:PDF}

\begin{figure}
  \centering
  \epsscale{1}
  \includegraphics{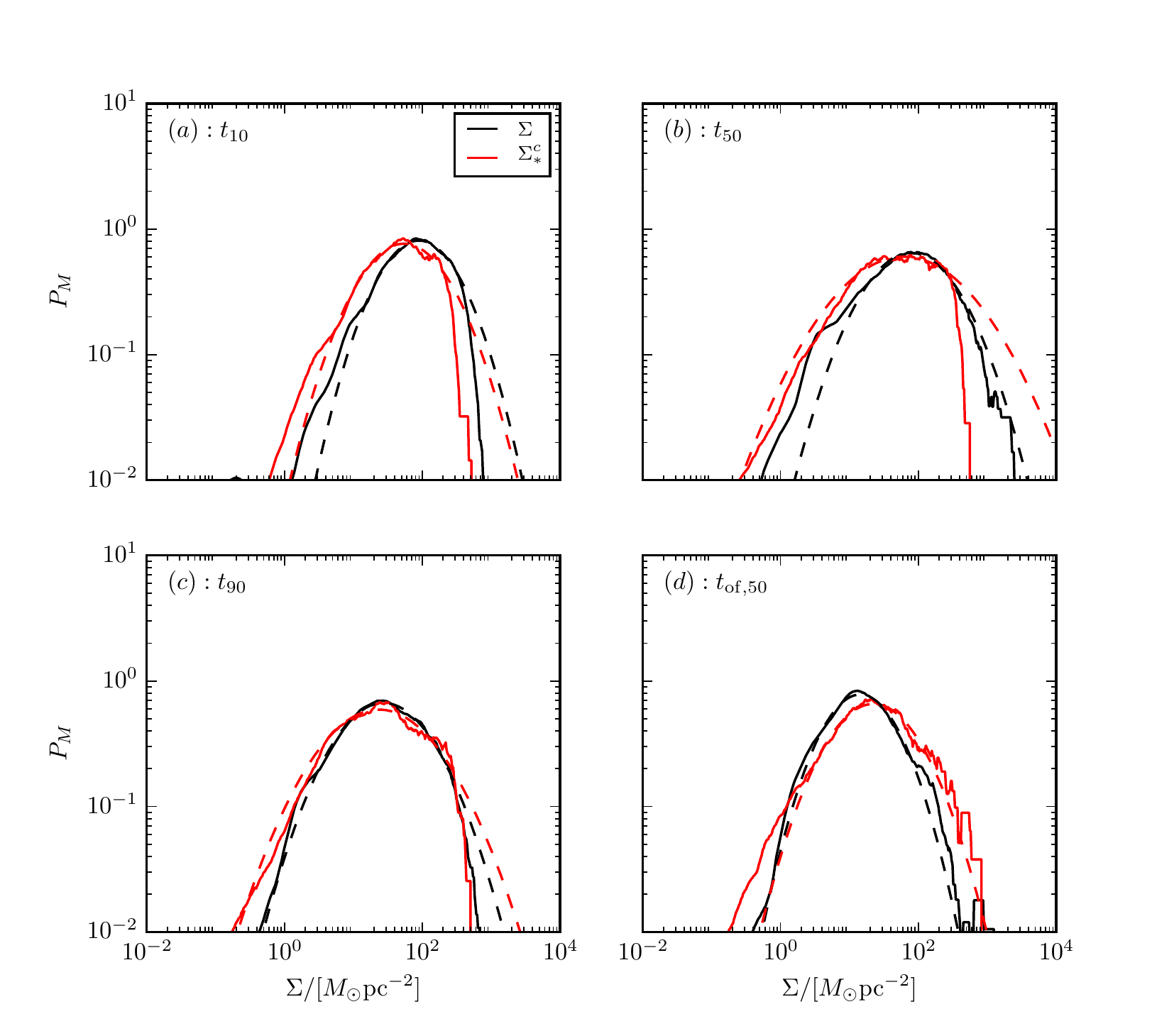}
  \caption{Surface density PDFs (by mass fraction) in the fiducial model at
    (a) $t = 0.56 t_{\rm ff} = t_{\rm 10}$, (b) $t=1.06 t_{\rm ff} = t_{\rm 50}$, 
    (c) $t=1.57 t_{\rm ff} = t_{\rm 90}$, and
    (d) $t=2.17 t_{\rm ff} = t_{\rm of, 50}$.  We show results for the observed surface density
    (black), and the circumcluster distribution beyond of the mean of
    the stellar distribution (red).  In each case, we show
    both the simulated surface density distributions (solid lines) as
    well as the best-fit lognormal curves (dashed lines).
    In fitting lognormals,
    the $15\%$ at low $\Sigma$ is omitted since that is
    approximately the fraction that is unbound by the initial
    turbulence and so is already outflowing from the cloud. The
    $10\%$ at high $\Sigma$ is also omitted, since the distribution is poorly
    sampled there.
\revis{For reference, with the opacity we adopt, surface density is related
    to optical depth by $\tau= 0.21 \Sigma/(\rm M_\odot \rm pc^{-2})$. With
    $\tau=1$ at $\Sigma =4.76\rm M_\odot \rm pc^{-2}$, well below the
    peaks in the mass PDFs, most of the mass is in structures that are
    quite optically thick to either external or internal FUV.}  
  }
  \label{Fig:LogNormalStar}
\end{figure}

\begin{figure}
  \centering
  \epsscale{1}
  \includegraphics{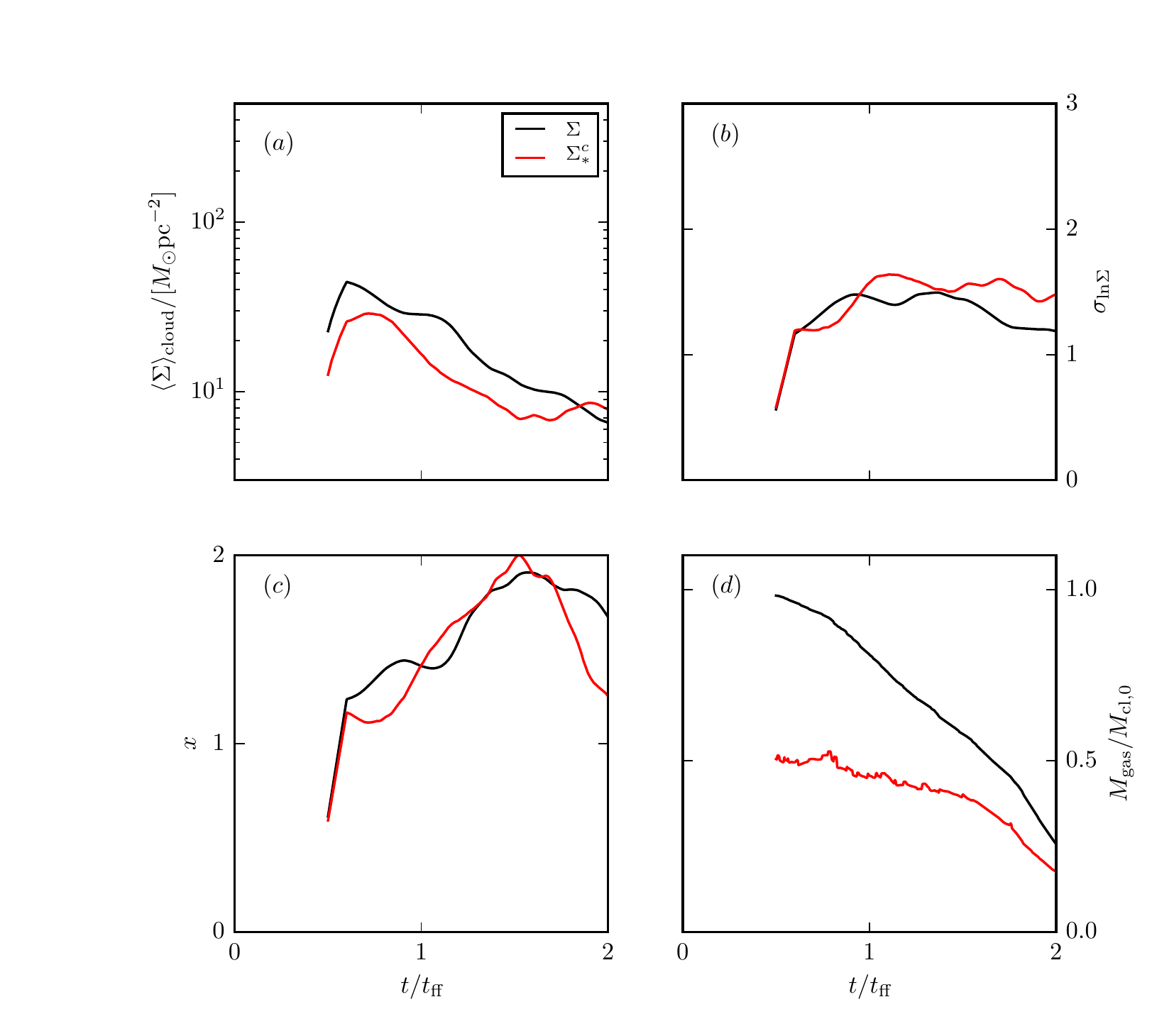}
  \caption{Best-fit lognormal parameters of surface density
    distributions in the fiducial model as a function of time. We show
    parameters for the fits for both  
    observed surface density ($\Sigma$) and circumcluster surface density 
    ($\Sigma_*^c$). In both cases, we show the best-fit 
    mean cloud surface density $\langle\Sigma\rangle_{\rm cloud}$ (top
    left), lognormal standard deviation $\sigma_{\rm \ln\Sigma}$ (top right),
    the mean cloud expansion
    factor $x$ (bottom left), and the total gas mass included in the
    fit (bottom right).}
  \label{Fig:LogNormalStarFit}
\end{figure}

We are most interested in the shape of the circumcluster surface
density distribution and how close this is to the lognormal form of
the observed surface density. In Figure~\ref{Fig:LogNormalStar}, we
show results from the fiducial model at four times: (a)
$t_{\rm 10}$, (b) $t_{\rm 50}$, (c) $t_{\rm 90}$, and (d)
$t_{\rm of,50}$.  At each time, we
show the PDF (mass fraction)
as functions of both observed surface density and
circumcluster surface density. The distributions of circumcluster 
surface density $\Sigma_{*}^c$ are
calculated by omitting gas inside the mean
stellar radius, since much of the gas
near the center of the cloud lies between stars, and hence would see
a radiation force from many surrounding star particles, which may
cancel out. Therefore, if we are interested in the interaction of gas
with radiation forces, the effective surface density distribution of
material surrounding the whole ensemble of stars is better
characterized by $\Sigma_{*}^c$. For any given circumcluster ray, the mass
is computed as $dM = d\Omega\int_{r_{\rm min}}^{r_{\rm max}} \rho r^2
dr$ where $d\Omega = d\phi \sin\theta d\theta$ is the solid angle.
In each panel we also show best-fit lognormals for each PDF.  

At all times both the observed and circumcluster mass PDFs appear to
be quite close to lognormal in shape.  The $\Sigma_{*}^c$ distributions are 
quite similar in their mean and variance to that of the ``observed'' $\Sigma$
distribution.  This suggests that the combination of turbulence and gravity
creates a structure so filamentary that there is little
difference between observing the cloud from outside and looking
outward from its center.


Even as the effects of radiation feedback become more pronounced, the
observed and circumcluster distributions remain similar, and both are
close to lognormal. Figure~\ref{Fig:LogNormalStarFit} plots best-fit lognormal parameters
for $P_M$ as a function of time for both the observed and circumcluster
PDFs for the fiducial model.
From the lognormal fit to each PDF, we obtain a mean
$\mu_M=\langle \ln\Sigma\rangle_M$ and standard deviation
$\sigma_{\rm \ln\Sigma}=\langle (\ln\Sigma -\mu_M)^2\rangle_M^{1/2}$.  
We define the ``cloud'' surface density via 
$\langle \Sigma \rangle_{\rm cloud}=
{\rm exp}(\mu_M - (1/2)\sigma_{\rm ln \Sigma}^2)$,
since this relation holds from the normalization of a lognormal.
The mean cloud surface density steadily
decreases with time (Figure~\ref{Fig:LogNormalStarFit}$a$),
while the best-fit variance remains approximately
constant throughout star formation
(Figure~\ref{Fig:LogNormalStarFit}$b$).

To obtain the ``cloud'' mass $M_{\rm cloud}$, excising low-density
ambient and outflow mass, we integrate the best-fit lognormal, shown
as a function of time in Figure~\ref{Fig:LogNormalStarFit}d.
The best-fit lognormal distribution to $P_M$ also gives us a way to
estimate the cloud's effective volume and size.
If we assume that the cloud is roughly
spherical, then we can define an effective cloud radius $r_{\rm eff} =
x r_0$ such that
\begin{equation}
	\langle \Sigma \rangle_{\rm cloud} \equiv \frac{M_{\rm cloud}}{\pi r_0^2 x^2}.
	\label{Eq:x}
\end{equation}
Figure~\ref{Fig:LogNormalStarFit}c shows the mean cloud expansion factor 
$x$ over time, based on this definition and the lognormal fit, which gives a
rough estimate of the effective cloud radius.

The character of the best-fit lognormal parameters appears to hold
true across the full range of simulations. Table~\ref{Tab:ModelParams}
shows $\sigma_{\rm ln \Sigma}$ and $x$ for the circumcluster surface density 
distribution at $t_{\rm 10}, t_{\rm 50}$, and $t_{\rm 90}$.  Almost
irrespective of surface density, $\sigma_{\rm ln \Sigma} \sim 1.4-1.6$
and $x \sim 1.2-1.6$. Furthermore,
the shape of $P_M(\Sigma)$ does not change much in time,
with $\sigma_{\rm ln \Sigma}$ varying by only around
$10-20\%$ from the beginning to the end of star formation.

We note that the mean values for $\sigma_{\rm ln \Sigma}$
in the present simulations are high compared to some 
estimates of observed molecular clouds. For example, \cite{Schneider2015}
measure the column density distributions for four GMCs in the Milky Way
(correcting for line-of sight contamination), and find
$\sigma_{\rm ln \Sigma}$ in the range $0.32-0.52$,
a factor $\sim 4$ smaller than in our models.
However, other current observational work suggests that variances
  may be nearly as large as those found in our simulations,
  with \citet{Lim2016} finding
  $\sigma_{\rm ln \Sigma} \sim 1.2-1.4$ in the IRDC G028.37+00.07, and emphasizing
  that results are sensitive to the
  detailed treatment of foreground and background corrections.
Magnetization tends to
limit compression by both turbulence and gravity,
and preliminary results from MHD simulations we have conducted
suggest that variances are
  somewhat lower than in the present (hydrodynamic-only) simulations.
In addition, the mean temperature in molecular gas may in fact be
closer to 20K 
than the conventional value of 10K which we have adopted
\citep{HeyerDame2015}; this would reduce $\sigma_{\rm ln \Sigma}$ by a few tenths.

With lower $\sigma_{\rm ln \Sigma}$, based on the theory presented in
Paper I the expectation is that SFEs would be reduced, but qualitatively
the interaction between radiation and gas would be similar to the results
we have found.
\revis{Lower $\sigma_{\rm ln \Sigma}$ would also imply that radiation could not
  escape from the cloud as easily.  If the escape fraction of radiation
  were lower, this would increase the momentum transferred to the gas.
  With cumulative absorption fraction $\sim 50\%$ for the present models
  (see Section \ref{SSS:escape}), this
  would be only a factor $\sim 2$ difference.  A reduced $\sigma_{\rm ln \Sigma}$
  would also reduce the variance in the velocity distribution of gas
  outflowing from the system, since as we shall show in Section
  \ref{sec:outflows} these quantities are directly related.}

\begin{figure}
  \centering
  \epsscale{1}
  \includegraphics{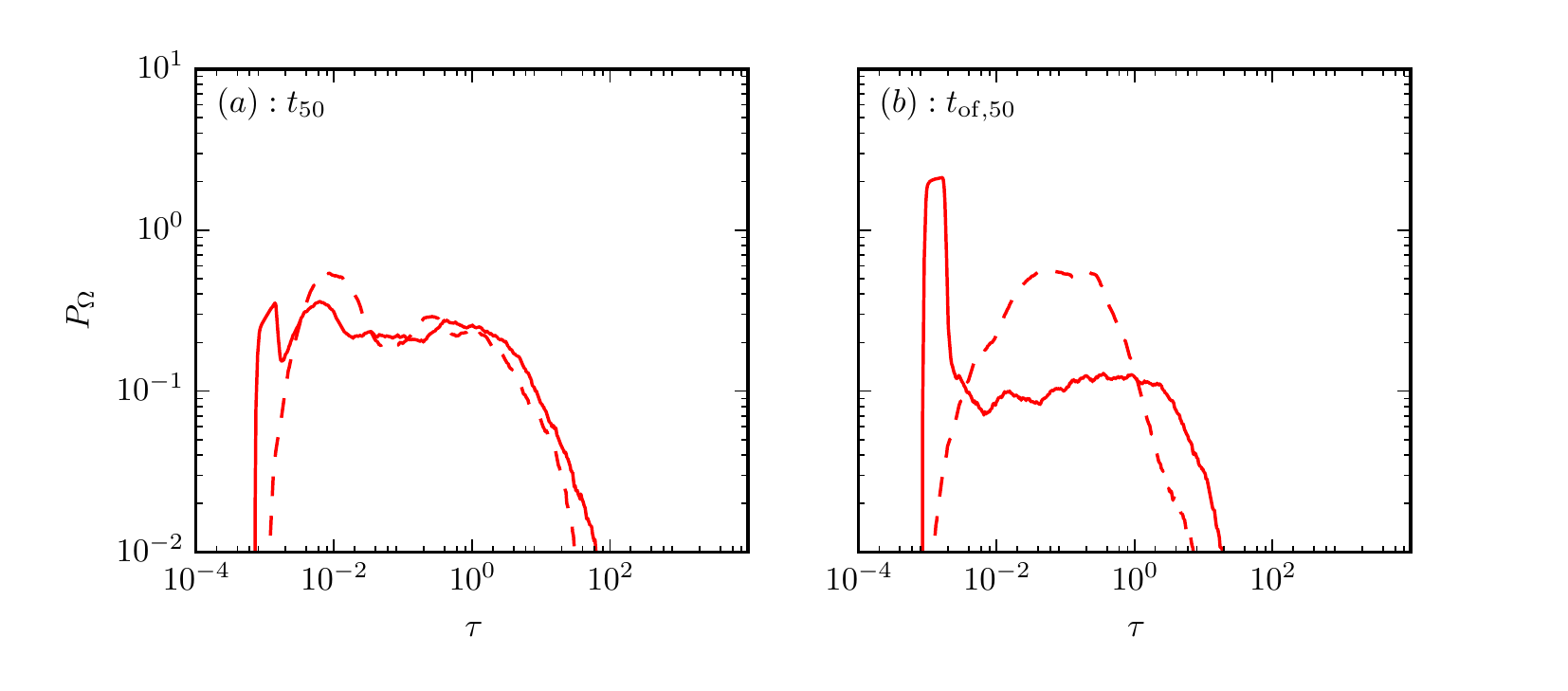}
  \caption{Optical depth PDFs (by solid angle fraction) at
    $t_{\rm 50}$ and $t_{\rm of, 50}$, for the fiducial model (solid curves), and
     a no-feedback model at the same times (dashed).
    We show results for 
    the circumcluster distribution outside the mean radius of the stellar
    distribution.
    \revis{While most of the solid angle looking outward
      from the center of the stellar distribution is at $\tau <1$,
      Figure \ref{Fig:LogNormalStar}
      shows that in contrast, most of the mass is in optically thick
      structures. 
      }
  }
  \label{Fig:Tau}
\end{figure}

For an externally-observed cloud, the mass distribution $P_M(\Sigma)$
is complemented by the area distribution $P_A(\Sigma)$.  However, in
considering circumcluster distributions, there is no single area that
characterizes any given ray because gas is at a range of distances.
Instead, we can consider the distribution in solid angle $\Omega$
with respect to the stellar center of mass.  In addition,
since this distribution is useful in characterizing how radiation is
absorbed, we consider $P_\Omega(\tau)$, where the optical depth $\tau$
is $\kappa \Sigma_*^c$.  Since we adopt a constant $\kappa = 1000 {\rm
  cm^2~g^{-1}}$, the PDFs in $\tau$ are linearly related to PDFs in $\Sigma_*^c$
using $\Sigma_*^c/(\rm M_\odot \rm pc^{-2}) = 4.76 \tau$.

In Figure~\ref{Fig:Tau} we show the optical depth distributions at
$t_{\rm 50}$ and $t_{\rm of, 50}$, for the fiducial model.
Unlike the distributions in mass, these distributions in solid angle
do not appear lognormal;  instead, both
during and towards the end of star formation, the PDFs are broad and
double-peaked, with the peaks separated by more than two decades in
$\tau$.  This would be unsurprising for distributions of area as a
function of optical depth (or of $\Sigma$) for the externally-observed case, 
because sightlines through the simulation domain sample both the cloud
and the low-density ambient medium, creating a bimodal distribution.
One might expect the ``ambient-only'' sightlines to disappear when
looking outward from the center of the cloud; instead, the
solid angle distributions of $\Sigma_*^c$ are even broader and flatter
than area distributions of ``observed'' $\Sigma$.

Comparison with $\Omega$ distributions for the no-feedback case,
also included in Figure~\ref{Fig:Tau}, show that the bimodal PDF
shape is not primarily a consequence of feedback.  Instead, it
reflects the fact that there are essentially two
optical depth distributions: one for sightlines that pass
through filaments and one for sightlines that do not.
Even without radiation forces clearing a path, there are natural channels in
a turbulent cloud through which radiation may escape.  We note that
\cite{Dale2012, Dale2013a} find a
similar result, namely that large holes in their clouds were present even when 
ionizing feedback was turned off in their simulations.  

Finally, we note that the effect of the large holes on $P_M$ is
minimal, because they contain very little gas by mass.  This also implies
that they will have little effect on the momentum of gas outflows, since
the same distribution is being accelerated outwards. However, the
holes have a large effect on the fraction of radiation escaping the cloud,
as we discuss next. 

\subsection{Escape of Radiation}
\label{SSS:escape}

The gas surrounding the central cluster 
is highly filamentary
with large holes that allow 
radiation to escape.
On the cloud or cluster scale, the amount of radiation
absorbed helps to regulate star formation.  For the present models,
in which radiation is the only regulation mechanism, 
a higher escape fraction implies a lower Eddington ratio (since
the total radiation force is $\sim (1-f_{\rm esc})L/c$),
which then requires a higher SFE and luminosity
to disrupt the cloud. In Paper I, we found SFEs a factor
of $10-20$ higher than would naively be expected by simple
considerations of the radiative force acting on a uniform shell.  
High escape fractions may partly explain this discrepancy.

Obtaining realistic estimates of radiation escape fractions from turbulent,
star-forming clouds is also important for understanding the ionization of
the diffuse ionized gas in the Milky Way
\citep[e.g.][]{HoopesWalterbos2000,VogesWalterbos2006}, as well as cosmic 
reionization from UV escaping dwarfs galaxies at high redshift
\citep[e.g.][]{Madau1999, Faucher-Giguere2008, Bunker2010,
  KuhlenFaucher-Giguere2012}.
Although the present models do not directly address radiative transfer
of ionizing radiation, the escape of non-ionizing UV is affected by
similar factors.

We may characterize how porous our clouds are to UV radiation by measuring the outward radiative flux
$\langle F_r \rangle_\Omega$ in spherical shells around the center of mass. With
$L_*$ the total stellar luminosity, the absorption fraction at radius $r$ is defined as 
\begin{equation}
	f_{\rm abs} (r) \equiv 1 - 4 \pi r^2 \langle F_r \rangle_\Omega / L_*,
\label{eq:fabsF}
\end{equation}
where $r$ is the radius of the sphere through which flux is computed.
We note that another measure of the absorption fraction is  
$f_{\rm abs, \tau} \equiv \langle 1 - {\rm e}^{-\tau} \rangle_\Omega$;
for our models this produces results that agree with $f_{\rm abs}$   
within $10$ or $15\%$

Until $\sim 1.5$~$t_{\rm ff}$, by which time the majority of star
formation is complete, we find
there is little difference between $f_{\rm abs}$ for models with and without
feedback. This confirms the results of Section~\ref{SSS:PDF}, i.e., that
direct radiation pressure has little effect on internal cloud
structure at early times, and is not effective at, for instance, driving the gas into
thinner shells or filaments. Therefore, as with ionized gas pressure
\citep{Dale2012, Dale2013a},
radiation pressure only modifies the radiation escape fraction by 
$\sim 5~\%$.

\begin{figure*}
  \centering
  \epsscale{1}
  \includegraphics{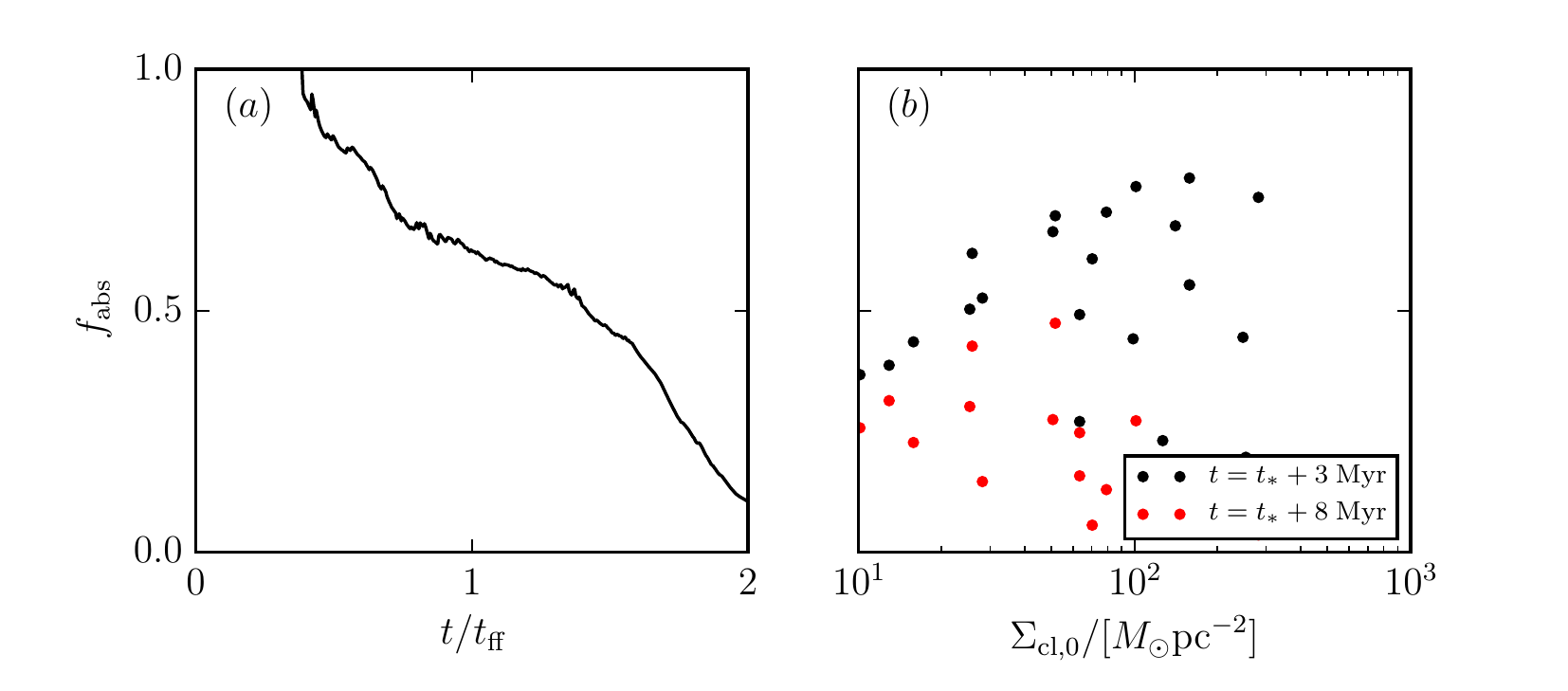}
  \caption{Absorption fraction, as defined in Equation (\ref{eq:fabsF}),
  measured at the box radius
  as a function of time for the fiducial model (left). We also
  show $f_{\rm abs}$ as a function of model surface density for all models
  (right).  Here, $f_{\rm abs}$ is measured 3 Myr (black) and 8 Myr (red)
    after star formation has begun in each model.}
  \label{Fig:FiducialAbs}
\end{figure*}

\begin{figure*}
  \centering
  \epsscale{1}
  \includegraphics{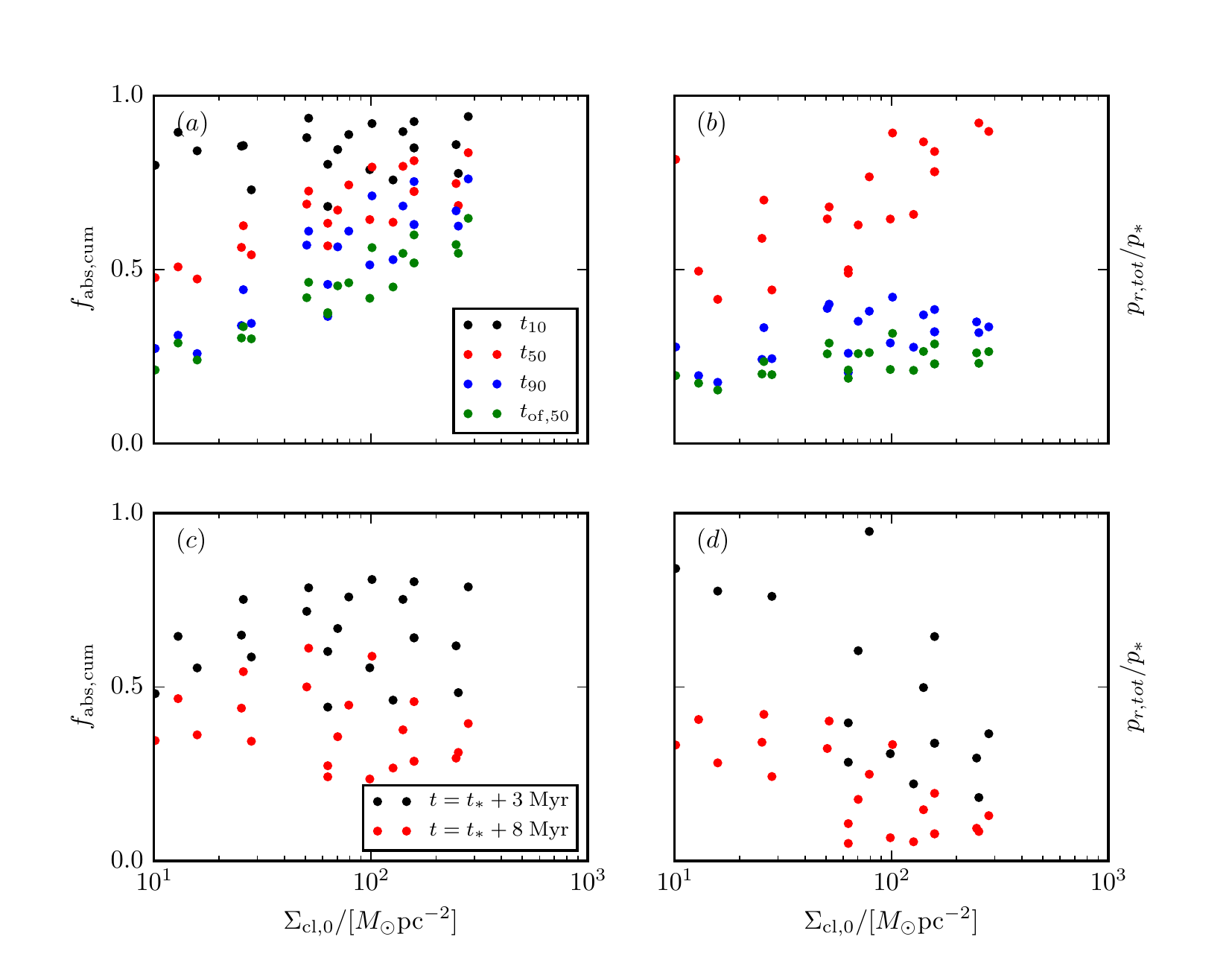}
  \caption{Cumulative absorption fraction (left) and ratio of the total radial 
  gas momentum $p_{r,tot}$ to total input stellar momentum $p_*$ (right), 
  each as a function of initial cloud surface density. In (a) and (b) 
  we show values at four times: when the SFE
  is at $10\%$ (black), $50\%$ (red) and $90~\%$ (blue) of its final value, and
    when half of the final mass has escaped
    the box (green). In (c) and (d), we show values for 3 Myr (black) and 8 Myr (red)
    after star formation has begun in each model.}
  \label{Fig:AbsCum}
\end{figure*}

Figure~\ref{Fig:FiducialAbs}$a$ shows the evolution of $f_{\rm abs}$ measured
at the box radius for the fiducial model, 
demonstrating that during the time when most of
the stellar growth occurs, $t/t_{\rm ff}\sim 0.7 - 1.6$, the
absorption fraction is relatively constant with roughly half of the radiation
escaping the cloud. Considering now the full set of models,
Figure~\ref{Fig:FiducialAbs}$b$ shows
the absorption fraction as a function of initial 
cloud surface density at two set times after the onset of star formation:
3 Myr, when the
first SNe could begin to explode, and 8 Myr, when the UV cluster
luminosity is expected to drop to half its original value due to the
loss of O stars \citep{Murray2005}.\footnote{Note, however, that the
  present simulations do not include SNe, nor do we allow the ratio of
  luminosity to mass to vary as a cluster ages.}
The values of $f_{\rm abs}$ at 3 Myr are also listed in Table~\ref{Tab:ModelParams}.

A useful quantity is the cumulative absorption fraction up to
a given time $t$, defined as the ratio of the total 
integrated flux absorbed to the total input stellar luminosity, and
computed via
\begin{equation}
	f_{\rm abs, cum} (t) = 1 - \frac{\int_0^t 4 \pi r^2 \langle F_r \rangle_\Omega dt}{\int_0^t L_* dt}.
\end{equation}
This quantity is plotted for several evolutionary stages and at two specific times
in Figure~\ref{Fig:AbsCum}.  We also show for comparison the ratio
of the total radial momentum of gas, $p_{r,tot}$,
to the total input momentum in stellar radiation, $p_* = \int dt L_*/c $. Here the 
total radial momentum of the gas
is defined as $p_{r,tot} = p_{r,box} + p_{r}$, which consists of 
the component contained within the simulation volume, 
$p_{r,box} = \int \rho {\bf v} \cdot {\bf \hat{r}} d^3x$, 
added to the time-integrated outflowing gas momentum 
$p_r = \int dt \int \rho ({\bf v} \cdot {\bf \hat{r}})^2 r^2 d\Omega$, both defined with reference to 
the stellar center of mass. Both the cumulative absorption fraction and the momentum 
ratio represent the efficiency of converting radiation momentum to gas momentum.

By definition, $f_{\rm abs,cum} < 1$, and from conservation of
momentum in an idealized spherical system (with zero initial
velocities), the value of $f_{\rm abs,cum}$ at late times should approach 
$p_{r,tot}  / p_*$.
Of course, our model
clouds are not ideal in that some flux is cancelled in the
decentralized stellar distribution and since a small amount of momentum is
also carried outward by the $\sim 12\%$ of gas mass that is
unbound by the initial turbulence.
More importantly, at earlier times, the initial turbulence
in the box can contribute non-negligibly to the measured
$p_{r,tot}  / p_*$.; this explains why some values of
$p_{r,tot}  / p_*$.
at 3 Myr (as listed in Table~\ref{Tab:ModelParams})
can exceed unity,
especially in low density clouds with correspondingly long freefall
times.

The values of the cumulative absorption fraction are on average higher than their instantaneous
counterpart, but the difference is only $\sim 50\%$. Thus, although most of the radiation is absorbed
at the earliest times, this stage is brief, and the luminosity is 
lowest then. We also note that the
trends with surface density remain the same. In particular,
irrespective of the initial cloud surface density, the cumulative
absorption fraction is $\sim 60~\%$ by $3$~Myr and $\sim 30~\%$ by
8~Myr.  At late times, the radial momentum of the gas is $\sim 25\%$ of
the level it would reach had the radiation field been spherically-symmetric
and perfectly absorbed (see Figure~\ref{Fig:AbsCum}b).

Our results on absorption fraction differ from those of \cite{Dale2012,
  Dale2013a} who find that for molecular clouds dominated by ionized
gas pressure, the absorption fraction generally increases with
surface density (offsetting higher SFE), and most cases therefore
contribute roughly the same ionizing luminosity 
to the ISM.  The difference does not likely to owe to 
differences in feedback, since neither our study nor theirs find much
change in internal cloud structure as a result of early feedback. Instead,
it may arise from the fact that they generally consider lower-surface-density 
clouds with freefall times much longer than 3 Myr,
hence they may not have yet been dispersed by that time. The one 
higher-surface-density cloud they model has an absorption fraction of only 
$\sim 10~\%$ by 3~Myr.

Finally, we note that the discrepancy between the absorbed and input
radiation is not enough to account for the high SFE of our simulated clouds. Simple models of cloud
destruction, which posit that star formation stops when the mean cloud
surface density reaches the Eddington surface density, would estimate 
for our fiducial model an efficiency of $\varepsilon \sim 0.03$, a factor
of $\sim 10$ smaller than the SFE obtained in our simulation.  An
extra factor of $2$ due to escaping radiation is therefore not
sufficient to explain why gas remains bound and continues to form
stars.
Instead, the reason for the high SFE is
that radiation momentum is not equally distributed to all of the mass
in the cloud.  Low-$\Sigma$ structures with large solid angles (per
unit mass) intercept much more radiation than required to reach
escape speed from the cloud, whereas high-$\Sigma$ structures continue
contributing  to star formation
until the luminosity and corresponding radiation flux are
finally sufficient to expel and unbind them.

\section{Gas Outflows}
\label{sec:outflows}

\begin{figure}
  \centering
  \epsscale{1}
  \includegraphics{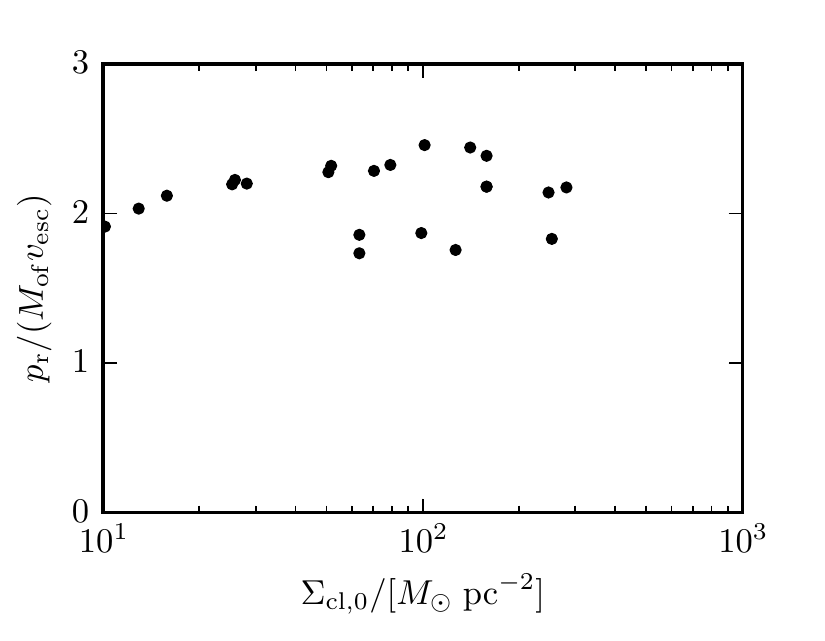}
  \caption{
    \revis{Mass-weighted mean}
    velocity of outflowing gas as a function of initial cloud surface density.
    For 
  comparison, we normalize to the initial escape speed of each model cloud.}
  \label{Fig:VSigmaVesc}
\end{figure}

We have established that the net radial momentum of the gas at late times
is significantly
less than the total radiation momentum input from stars,
owing to the relatively high radiation escape
fraction (Figure~\ref{Fig:AbsCum}$b$).
Figure~\ref{Fig:VSigmaVesc} directly shows the ratio of outflowing gas momentum
to outflowing gas mass, in units of the escape speed at the cloud's
initial radius (both momentum and mass outflows
are integrated over the duration of the simulation).  
Evidently, the outflowing gas is unbound, and in fact the
(scaled) mean velocity of outflowing gas is relatively constant
across our full range of simulations, in the range $p_r/(M_{\rm of} v_{\rm esc})
\sim 1.5-2.5$ for all models.  However, of equal interest is the
distribution of momenta and velocities for the escaping gas; i.e.
to what extent does 
radiation pressure drive fast low-density outflows vs. slow high-density 
outflows in a given system?

Characterizing the properties of outflows driven by radiation pressure
is particularly interesting in connection to the origin of ``cool,'' fast
winds from starburst galaxies
({note that unlike in the ISM,
  in this context ``cool''
  means $T<10^6$K}).  
These winds emerge at velocities of
up to $\sim 1000~{\rm km~s^{-1}}$, and are seen in gas that is cool
enough to show absorption in NaD or MgII.
The origin of this
high-velocity cool material is not understood \citep{Veilleux2005}.  
Possibilities include entrainment of dense clouds 
\citep[e.g.,][]{2015ApJ...805..158S}
by the outflowing hot gas created by supernova explosions
\citep[e.g.,][]{ChevalierClegg1985}, or radiative cooling of 
hot outflows that are themselves sufficiently dense 
\citep[e.g.,][]{Wang1995,Thompson2016}, or acceleration by radiation pressure
forces acting on cloudlets exposed to the combined radiation field 
of thousands of luminous stars created in a 
starburst \citep[e.g.,][]{Murray2005,Thompson2015}. Although the
present simulations do not study clusters as massive as those powering
starbursts, they do provide valuable insight into the driving of winds
by radiation pressure, as the interaction of radiation and gas is implemented
here via fully self-consistent RHD. 

\citet{ThompsonKrumholz2016} have pointed out that when considering
wind driving by radiation pressure, the lognormal distribution of
density is important because structures of low surface density may be
super-Eddington even when the circumcluster gas distribution as a
whole is not.  In our simulations, as the luminosity increases over
time, an increasing fraction of the gas becomes super-Eddington and is
driven out of the cloud, until the last remnants are swept clean.
\citet{Thompson2015} argue that the asymptotic velocities of
structures accelerated by a luminous source will depend on both the
source luminosity and the distance at which they become optically thin,
and that for certain parameter regimes, velocities can be much greater 
than the escape speed of the system. 
We can use our RHD simulations to test these and other ideas related
to the properties and statistics of radiation-driven winds from
turbulent, cold, star-forming gaseous systems.

\begin{figure}
  \centering
  \epsscale{1}
  \includegraphics{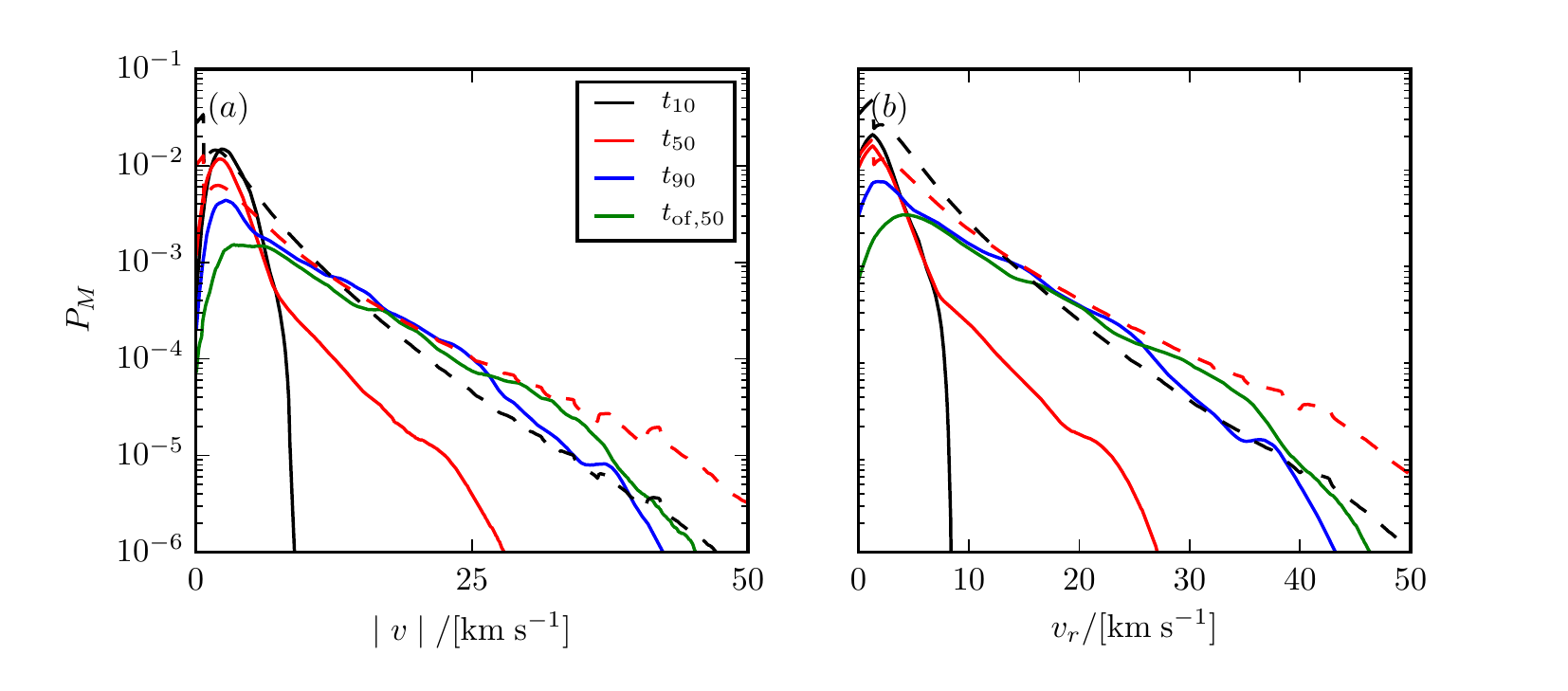}
  \caption{Probability density functions of the velocity in our
    fiducial simulation for both the absolute velocity (left) and the
    radial velocity away from the center of mass (right). We show
    results at $t/t_{\rm ff} = 0.56, 1.06, 1.57$, and $2.17$, corresponding
    to $t_{\rm 10}, t_{\rm 50}$,  
    $t_{\rm 90}$, and $t_{\rm of,50}$.  We show also theoretical velocity
    distributions obtained by applying Equation~(\ref{Eq:Pdv}) with a best
    fit lognormal distribution over surface density to two assumed
    density profiles: uniform (black dashed) and
    $r^{-1}$ (red dashed), respectively}
  \label{Fig:FiducialV}
\end{figure}

\begin{figure}
  \centering
  \epsscale{1}
  \includegraphics{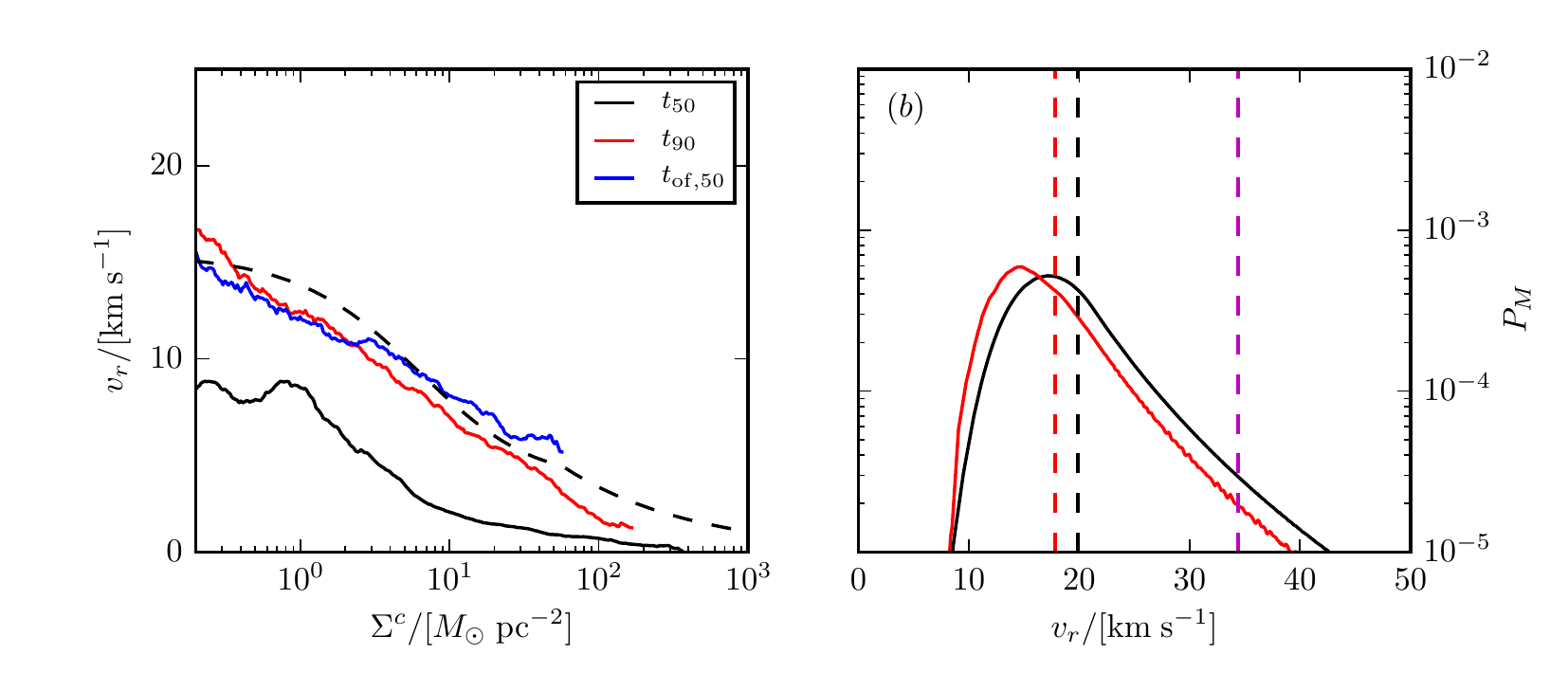}
  \caption{{\it Left:} Mean radial outflow velocity as a function of circumcluster surface
    density at different stages in the evolution of the fiducial model, 
    calculated as the total momentum along a radial sightline divided
    by the total gas mass. We overplot the expected
    velocity (black dashed line) calculated using Equation~(\ref{Eq:vrf}) with 
    $M_* = \varepsilon_{\rm final} M_{\rm cl, 0} = 0.42 M_{\rm cl, 0}$; $r_0 = r_{\rm cl,0}$; and 
    $r = 2r_{\rm cl,0}$. We omit $\Sigma^c < 0.1$ since these surface densities consist of 
    primarily the lowest density gas which has generally dropped below the 
    computational density floor and was reset to a minimum density and zero velocity. 
    {\it Right} shows the mass distribution as a function of asymptotic outflow velocity, using
    Equation~(\ref{Eq:Pdv}) with both a best-fit lognormal surface
    density distribution (black) and the full simulated surface
    density distribution when $t=0.57~t_{\rm ff}$ (red), with
    $\rho(r) \propto r^{-1}$ and 
    only considering surface densities such that $\Sigma^c < \Sigma_{\rm E,max}$.
    Dashed vertical lines show the mean velocity in each case,
    as well as the velocity expected for expansion of a uniform shell (magenta).}
  \label{Fig:FiducialVSigma}
\end{figure}

We consider first the velocity distribution of gas both in the cloud
and escaping it. In Figure~\ref{Fig:FiducialV} we show the PDF of mass
as a function of velocity, both absolute and radial, for gas in the
fiducial model at various stages of evolution. At early times, the
radial velocity distribution is very close to the Gaussian
distribution of our initial turbulent field, with slight excesses at
negative velocities corresponding to gas accreting on to the central
star and at positive velocities corresponding to gas being driven
from the cloud. However, once star formation feedback becomes
significant, the distribution of velocities above the escape velocity
of the cloud, $v_{\rm esc} \sim 5.4~{\rm km~s^{-1}}$, departs
dramatically from this Gaussian shape.


To understand the distribution of high velocity gas, we first consider
the relationship between velocity and surface density on individual
rays emerging from the stellar center of mass. In
Figure~\ref{Fig:FiducialVSigma}$a$, we show the mean
mass-weighted velocity along sightlines as a function of their surface density. 
The radial velocity for a given surface density bin is calculated by
computing the net radial momentum of gas along all rays within that bin, 
then dividing by the total mass along those rays. 


The relationship between surface density and velocity can be understood by
considering individual structures interacting with a central source of
radiation and gravity, becoming unbound when their Eddington factors
exceed unity.  We follow the motion of a structure initialized 
(from rest) at radius $r_0$ with surface density $\Sigma_0$.  We
assume the structure remains within a fixed solid angle with respect to
the source, so that as it moves outward, its surface density decreases
with distance $r$ as $\Sigma=\Sigma_0 (r_0/r)^2$
({note that we simply use $\Sigma$ for this
  idealized ballistic calculation, reserving $\Sigma^c$ for circumcluster
surface densities measured in our simulations}).  The inward gravitational
acceleration from the star cluster is $GM_*/r^2$, while the outward radiation
force per unit mass is $\Psi M_* (1-\exp(-\tau))/(4\pi c \Sigma_0 r_0^2)$.
Here, we have used $L_*=\Psi M_*$ for the cluster luminosity, and the
optical depth of the structure is $\tau = \kappa \Sigma = \tau_0 (r_0/r)^2$
for $\tau_0=\kappa \Sigma_0$.

The equation of motion then becomes 
\begin{equation}
	\frac{dv}{dt} = \frac{\Psi M_*}{4 \pi c \Sigma_0 r_0^2}\left(1 - {\rm e}^{-\tau}\right) - 
	\frac{G M_*}{r^2}
	= \frac{G M_*}{r_0^2}\left[\frac{\Sigma_{\rm E,max}}{\Sigma_0}
          \left(1 - {\rm e}^{-\tau_0 (r_0/r)^2)}\right) 
	- \left(\frac{r_0}{r}\right)^2\right],
	\label{Eq:EOM}
\end{equation}
where we have introduced a maximum Eddington surface density
\begin{equation}
  \Sigma_{\rm E,max} \equiv \frac{\Psi}{4 \pi c G}
  = 380~{\rm M_{\odot}~pc^{-2}} 
	\left(\frac{\Psi}{2000~{\rm erg~s^{-1}~g^{-1}}}\right).
\label{eq:SigmaEmax}
\end{equation}
The quantity $\Sigma_{\rm E,max}$ 
represents the largest surface density for which the stellar Eddington ratio
\begin{equation}
	f_{\rm Edd,\star} \equiv \frac{F_{\rm rad}}{F_{\rm grav,\star}} = 
	\frac{\frac{\Psi M_*}{4 \pi c r^2}\left(1 - {\rm e}^{-\tau}\right)}{\frac{G M_*}{r^2}\Sigma}=\frac{\Sigma_{\rm E,max}}{\Sigma}(1-e^{-\tau})
        \label{eq:fedd*}
\end{equation}
can be larger than unity, so that the radiation force overcomes
gravity. Thus, only structures with $\Sigma < \Sigma_{\rm E,max}$ can be
accelerated outwards. We note that in reality, an accelerating fluid element will feel the
gravity of the full remaining cloud mass, not just the central star
cluster. However, once most of the cloud mass has been
expelled, this is a small correction. In fact, at the point 
the fluid element becomes unbound and its radius
begins to increase, even the stellar gravity term becomes negligible
compared to the radiation term.
To see this, note that while the structure remains optically thick, the
radiation force in Equation (\ref{Eq:EOM}) is constant, while gravity
$\propto r^{-2}$.  If the structure becomes optically thin,
taking $\tau \ll 1$ in Equation (\ref{eq:fedd*}) shows that 
the Eddington ratio approaches a maximum value given by 
\begin{equation}
f_{\rm Edd,*, max}=\kappa \Sigma_{\rm E,max}  \equiv \tau_{\rm E,max} = 80
  \left(\frac{\Psi}{2000~{\rm erg~s^{-1}~g^{-1}}}\right)
  \left(\frac{\kappa}{1000{\rm ~cm^2 ~g^{-1} }}\right),
\end{equation}
which is large for our fiducial parameters.
Provided that star formation is slow compared to the time to accelerate
gas out of the cloud (i.e. provided  the acceleration timescale is short compared
to the free-fall time), we can also treat $M_*$ as a constant in
Equation (\ref{Eq:EOM}). 

Multiplying Equation (\ref{Eq:EOM}) by $v=dr/dt$, we integrate over $t$ to
obtain the velocity as a function of distance $r$, given by 
\begin{equation}
  \frac{  v^2(\Sigma_0,r)}{v_{\rm esc}^2(r_0)} =
  \frac{\tau_{\rm E,max}}{\tau_0}
  \left\{ \sqrt{\tau_0 \pi}
    \left[{\rm erf}\left(\sqrt{\tau_0}\right) -
      {\rm erf}\left(\frac{\sqrt{\tau_0}}{r / r_0}\right)\right]
    +\frac{r}{r_0}\left[1 - {\rm e}^{-\tau_0\left(r_0/r\right)^2}\right] 
	+ {\rm e}^{-\tau_0} - 1\right\} + \frac{r_0}{r} -1 .
	 \label{Eq:vrf}
\end{equation}
Here, we include only the star cluster mass in
$v_{\rm esc}(r_0)\equiv (2 G M_*/r_0)^{1/2}$.
Note that Equation (\ref{Eq:vrf}) may also be used to write the velocity
in terms of the density $\Sigma$ at a distance $r$ by
substituting $\tau_0= \kappa \Sigma r^2/r_0^2$.  

At large distances from the source $r/r_0 \gg 1$, 
\begin{equation}
  v^2(\Sigma_0) \rightarrow v_{\rm esc}^2(r_0)
    \left\{
 \frac{\tau_{\rm E,max}}{\tau_0}
  \left[ \sqrt{\tau_0 \pi}~
 {\rm erf}\left(\sqrt{\tau_0}\right) + {\rm e}^{-\tau_0} - 1\right]  -1\right\} .
\label{Eq:vrasy}
\end{equation}
Therefore, structures that start as optically thin ($\tau_0 \ll 1$) have
$v/ v_{\rm esc} \rightarrow (\tau_{\rm E, max} -1)^{1/2}\approx 9$,
while structures that
start as optically thick ($\tau_0 \gg 1$) eventually become optically thin at
$r/r_0=\tau_0^{1/2}$ and reach a final speed 
$v/ v_{\rm esc}\rightarrow [\tau_{\rm E, max}(\pi/\tau_0)^{1/2}-1]^{1/2}$.
That is, structures that are initially optically thin accelerate to a velocity
nearly ten times the escape speed at their launch point,
while this is reduced by a factor $\sim (\pi/\tau_0)^{1/4}$
for  structures that are initially optically thick.
For our parameter choice $\kappa=1000 {\rm ~g~ cm^{-2}}$, the $\tau=1$
transition between optically thick and thin occurs at a surface density of 
$\Sigma_{\rm th} =\kappa^{-1}= 4.8~M_{\odot}~{\rm pc^{-2}}$
(or hydrogen column
$N_{\rm th} = 4.3\times 10^{20} ~{\rm cm}^{-2}$).
Because optical
depth decreases $\propto r^{-2}$, the asymptotic velocity in the optically-thick
limit is essentially the same as obtained by taking the optically-thin limit
for a launch radius where the optical depth is equal to $\pi$.  We note
that Equation~(\ref{eq:SigmaEmax}) imposes a limit $\tau_0 < \tau_{\rm E,max}$
on the optical depth of structures that can be accelerated at all,
implying a lower limit on the velocity of
escaping structures at large distance:
$v/v_{\rm esc}(r_0) > [(\pi \tau_{\rm E,max})^{1/2} -2]^{1/2}\sim 3.7 $.  Thus, there
is a range of only about a factor of two between the minimum and
maximum asymptotic velocities for structures originating at a given radius $r_0$.  
However, structures that begin with optical depth near
$\tau_{\rm E, max}$ would require an order of magnitude increase in radius to 
reach their asymptotic velocity. In practice, this asymptotic velocity
may not be reached unless the cluster and cloud are relatively isolated
within their enviroment.  

For our simulations, acceleration starts at
$r_0\lsim r_{\rm cl,0}$ and continues until the gas leaves the box at $r \sim
2r_{\rm cl,0}$. We may therefore compare expectations of radiative acceleration
to our simulation results by plotting
Equation~(\ref{Eq:vrf}) for $r = 2r_{\rm cl,0}$, $r_0 = r_{\rm cl,0}$ and 
$M_* = 0.42M_{\rm cl,0}$ (corresponding to the final stellar efficiency of the fiducial 
model).  Here, we take
care to adjust for the fact that by the time fluid elements reach 
the edge of the box, the circumcluster
surface density $\Sigma^c \sim \Sigma_0 / 4$ (we also omit the ``*'' subscript
in $\Sigma_*^c$
since at late time
the circumcluster material is well outside the star particle distribution).
The resulting predicted 
relationship between $v$ and $\Sigma^c$ is shown in
Figure~\ref{Fig:FiducialVSigma}, compared to the measured 
relationship between mean velocity and $\Sigma^c=\int\rho~dr$ obtained
along lines of sight in the simulation at various stages.
We see that Equation~(\ref{Eq:vrf}) captures the late-time relationship between
velocity and surface density well.  The match is poor at early times,
but this is as expected since our plotted data use the final stellar mass and 
the gas has not yet had a chance to expand to the
edge of the box.
Notably, the simulation results are generally consistent with 
the predicted power-law relation between $v$ and $\Sigma^c$ at large surface
density (corresponding to the $v\propto (\Sigma^c)^{-1/4}$ behavior 
expected in the optically thick limit).

For comparison to the distribution of velocity with mass shown in
Figure~(\ref{Fig:FiducialV}), we consider the full range of
launch points $r_0$ and the PDF of $\Sigma^c$ by mass at
each launch point.  A structure of given $\Sigma_0$ will accelerate to
higher asymptotic velocity if it is launched from $r_0$ nearer to the
cluster center.  Conversely, a fluid element starting with low
$\Sigma_0$ and large $r_0$ may be able to reach the same asymptotic
velocity as a fluid element starting from high $\Sigma_0$ at small
$r_0$.  In particular, starting from optically-thick conditions (as
is true for most of material in the cloud),
Equation (\ref{Eq:vrasy}) shows that the
asymptotic velocity scales as $v \propto r_0^{-1/2}\Sigma_0^{-1/4}M_*^{1/2}$,
i.e., inversely with the fourth root of gas mass per unit solid angle
(and directly with the square root of the cluster mass).
Thus, to obtain $P_M(v)$, we must consider the fraction of mass at
each radius in the cloud that is in structures of a given surface
density, and integrate over the distribution of mass with radius.

For the distribution of mass with radius, we adopt the spherically-symmetric 
power-law density profile in radius $\rho(r) \propto
r^{-\alpha}$ discussed in Section~\ref{SSS:profile}. The distribution
of possible starting radii is then $P_M(r_0)dr_0 = dM / M \propto
r_0^{2-\alpha}dr_0$, which leads to a mass-weighted mean value of inverse radius equal
to $r_{\rm cl}^{-1}(3-\alpha)/(2\alpha)$.
Equation~(\ref{Eq:vrf}), which represents $P_M(v |
\Sigma_0, r_0)$, may then be used to derive the theoretical 
distribution of outflow velocity by marginalizing over our initial lognormal
distribution for $\Sigma_0$ and the assumed distribution for $P_M(r_0)$ to obtain
\begin{equation}
  P_M(v)dv = \int dr_0 \int d\Sigma_0
  P_M(v | \Sigma_0, r_0) P_M(r_0) P_M(\Sigma_0) dv,
	\label{Eq:Pdv}
\end{equation}
where we have assumed the independence of $\Sigma_0$ and $r_0$.
The resulting outflow velocity distribution (taking $r = 2r_{\rm cl,0}$),
shown in
Figure~\ref{Fig:FiducialV} for $\alpha = 0$ (uniform density
distribution) and $\alpha = 1.0$ (best-fit to our simulations),
compares remarkably well to the final velocity distribution. 
Necessarily, we do not capture the distribution below
$v_{\rm esc}$ since we are only considering outflows. Furthermore, we
overestimate the mass in the high-velocity tail since the
theoretical distribution, which uses the surface density distribution at 
the onset of star formation, does not account for mass leaving the
simulation volume. This becomes quite pronounced by the time half the mass has left the
box. However, at velocities below this tail (and above $v_{\rm esc}$) 
the theoretical prediction captures the shape of the
distribution obtained in the simulation rather accurately.

Given how well the simple theoretical model matches the measured velocity
distributions in the ``near'' region (i.e., within our computational domain), 
it is interesting to apply it to predict asymptotic velocities at
large distance.  Adopting a density profile $\rho \propto r^{-1}$
within the cloud, Figure~\ref{Fig:FiducialVSigma}b shows what the
distribution would become if all gas were accelerated until it becomes
optically thin, using Equation~(\ref{Eq:Pdv}) and the fiducial model's 
surface density distribution $P_M(\Sigma^c)$ at $t_{10}$. 
Around $99~\%$ of the mass in the fiducial model is below a velocity of
$\sim 40~{\rm km~s^{-1}}$
or close to $8$ times the escape velocity from the original edge of the cloud.
Although the PDFs of cloud material have a slight excess above a lognormal
at low $\Sigma$ (e.g., see Figure~\ref{Fig:LogNormalStar}), 
this does not result in any significant
difference in the outflow velocity distribution because it represents 
very little mass.
The peak of the distribution in Figure~\ref{Fig:FiducialVSigma}b
has a roughly lognormal shape because the relationship between asymptotic
velocity and surface density is $v \propto \Sigma_0^{-1/4}$ for optically
thick structures (most of the material), and the underlying $P_M(\Sigma_0)$
is lognormal.  

In our simulations, most of the gas mass is in structures that are
optically thick to UV, i.e., with $\Sigma^c > \Sigma_{\rm th}=\kappa^{-1} \sim
4.8 M_\odot {\rm pc}^{-2}$. This is also true by definition in real
molecular clouds, because at lower column, molecules would be dissociated
by UV.  Only the small amount of gas that begins in optically thin structures 
will reach velocities as large as $\sim 9 v_{\rm esc}(r_0)$.
On the contrary, most of the gas begins in optically thick structures, becoming optically
thin only at distances where $v_{\rm esc}(r_0)$ has dropped below
that of the original cloud (i.e., for $r>r_{\rm cl,0}$),
and would therefore have lower asymptotic ``wind'' velocity. 
Due to the lognormal distribution, with most of
the mass at surface densities well above the mean value for the cloud,
these structures would accelerate more slowly and reach lower final speeds
than would be true for a uniform-density cloud.  
For the fiducial model, accounting for the lognormal distribution,
the predicted mean outflow velocity would be less than
$20~{\rm km~s^{-1}}$  (Figure~\ref{Fig:FiducialVSigma}b).  By comparison, if an 
optically-thick shell of mass $(1-\varepsilon)M_{\rm cl,0}$ uniformly-distributed 
in solid angle were accelerated outward by radiation from a
cluster of mass $\varepsilon M_{\rm cl,0}$, it would reach an asymptotic velocity
(see Equation~\ref{Eq:vrasy}) given by 
\begin{equation}
  v_{\rm unif} \rightarrow \left(\frac{\Psi \varepsilon}{c}\right)^{1/2}
  \left(\frac{\kappa M_{\rm cl,0}}{1-\varepsilon}\right)^{1/4},
\label{Eq:vruni}
\end{equation}
which would be $\sim 35~{\rm km~s^{-1}}$ for the fiducial model.  While
these overall scalings with cloud mass and SFE 
are still expected to apply
\footnote{
  From Equation~(\ref{eq:SigmaEmax}),
  $\Sigma^c<   \Sigma_{\rm E,max}$ is
  required for ejection of any given structure, and if the final SFE 
  is controlled by radiation, this implies a value for   
  $(1-\varepsilon)M_{\rm cl,0}/r_0^2$ that depends in detail on
  the lognormal distribution (Paper I).  In terms of overall scalings,
  this leads to $M_{\rm cl, 0}^{1/4}/(1-\varepsilon)^{1/4}\propto v_{\rm esc} (c/G \Psi)^{1/4}$.   
},
the lognormal density distribution in
the cloud tends to reduce the mean value of the ``wind'' velocity,
while broadening the overall distribution.
Even though velocities are lower than the naive prediction, radiation is still capable of driving
outflows at a mean final velocity of $3-4$ times the
initial escape velocity from the cloud's surface.
Note that this is roughly twice as large as the mean value 
$p_r/(M_{\rm of} v_{\rm esc}) \sim 2$ shown in
Figure~\ref{Fig:VSigmaVesc} as measured from our simulations, because
acceleration would continue outside of our simulation domain.

\begin{figure}
  \centering
  \epsscale{1}
  \includegraphics{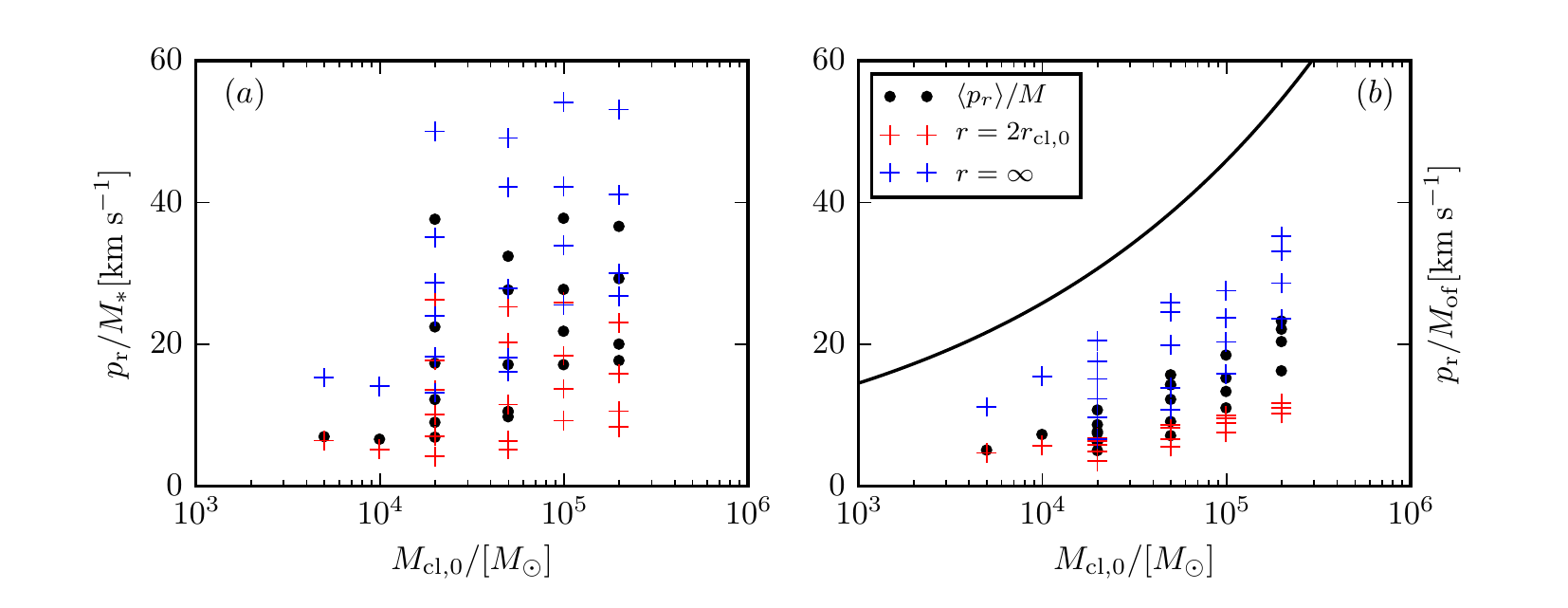}
  \caption{Outflowing gas momentum generated by direct radiation pressure,
    per unit stellar mass formed (left) and per unit outflowing mass (right), 
    as a function of initial cloud mass. Results from all models are shown
    (black dots). For comparison, we also show
    the theoretical mean velocities (crosses) based on Equations~(\ref{Eq:vrf})
and (\ref{Eq:Pdv}) 
    at $r = 2r_{\rm cl,0}$ (red) and for $r
    \rightarrow \infty$ (blue), as well as the equivalent final velocity for 
    a uniform shell from Equation~(\ref{Eq:vruni}), with $\varepsilon=0.5$ (black line)}.
  \label{Fig:VInfSigma}
\end{figure}

Our conclusion that non-uniform surface density systematically lowers
the mean outflowing gas velocity holds true across the
cloud mass and radius variations explored in our simulations.  In
Figure~\ref{Fig:VInfSigma} we show the mean radially-outflowing cloud
momentum per unit outflow mass (i.e., the mean outflowing gas velocity)
and per unit stellar mass, as functions of the initial cloud mass. In
all cases, the theoretical outflow velocity predicted by Equation~(\ref{Eq:vrf}),
when integrated over the best-fit lognormal surface density distribution
and power-law density profile via Equation~(\ref{Eq:Pdv}),
does a reasonable job of capturing the
simulated momentum per unit outflow mass.
The estimate is slightly less accurate at higher cloud masses, due
to the effect of outflows caused by turbulence in the
initial conditions (this causes more bias in high-mass clouds, because
the radiation-driven outflows are a smaller fraction of the total).

Finally, we note that based on the statistical properties
of our simulations and the radiation acceleration
model, the value of the outflow momentum in gas per stellar
mass formed, $p_r/M_*$, is expected to be in the range
$20-80~{\rm km~s^{-1}}$ 
(the model includes acceleration beyond the simulation box, and therefore
yields values larger than reported in Table~\ref{Tab:ModelParams}).
This is far below the simple estimates of the momentum injection by radiation
forces often adopted in the literature, which use $p_r/M_*
\sim 180~{\rm km~s^{-1}}$  based on the assumption that all the radiation
over the lifetime of a cluster is absorbed.  There are several factors
that contribute to this reduction: (a) a large fraction of the
radiation escapes due to the highly inhomogeneous gas
distribution created by turbulence, (b) the increase of the
mass-weighted mean surface density (again due to turbulence) reduces
the acceleration of fluid elements, and (c) any individual gas structure
becomes optically thin as it expands away from the cluster, thus limiting
the total radiation it absorbs.  The low values of $p_r/M_*$ from this process compared to
that for radiative supernova remnants ($\sim 3000~{\rm km~s^{-1}}$; e.g.
\citealt{KimOstriker2015}) implies that
momentum injection by UV radiation is not important to driving
turbulence or otherwise providing dynamical support to the bulk of the
ISM in galaxies.  Nevertheless, UV radiation 
may still play an important role in accelerating a small fraction of the gas
to high velocities ($\simge 100\ {\rm km/s}$) in starbursts,
as Equations (\ref{Eq:vrasy})
and (\ref{Eq:vruni}) suggest for regions with a large mass
($\sim 10^7-10^8 M_\odot$) in luminous
stars and a high escape speed.

\section{Summary and Discussion}
\label{Sec:Conclusion}

In Paper I, we presented the first results from a set of simulations
that study gravitational collapse, star formation, and gas dispersal
by UV radiation forces in turbulent GMCs.  There, we
characterized the distributions of the gas column density produced by
turbulence as log-normal PDFs, analyzed the dependence of the SFR and
final SFE on cloud properties, and related the final SFE in each cloud
to the properties of the PDF via the condition that the outward
radiation force must exceed the inward gravitational force for
individual dense structures to be ejected.  Here we further consider
aspects of cloud structure produced by turbulence combined with
self-gravity, and investigate the implications of this structure for
the interaction with radiation from embedded star clusters.  We use
our simulations to quantify the escape of stellar UV radiation from
filamentary clouds in which star clusters are born, to measure the
total momentum transfer from radiation to gas, and to examine the detailed
distribution of mass with velocity in the outflowing gas.

Our main conclusions are as follows:

\begin{itemize}
  
\item[1.]{\it Cloud Evolution and Star Clusters}

The clouds in our simulations are highly inhomogeneous throughout their
evolution, with most of the mass concentrated in dense filaments and
star formation exclusively within these structures.  Radiation emerging
from embedded clusters removes the lower density gas from the cloud,
creating large holes through which radiation can escape.  At early times
(until at least halfway through the star formation process), the overall
density profile of gas in the cloud is fairly shallow, having $\rho \propto r^{-\alpha}$
with
$\alpha \sim 0.8-1.3$ (see Figure~\ref{Fig:RhoProfile}).
Since all models in this work are initialized with the same virial
parameter ($\alpha_{\rm vir} = 2)$, corresponding to a marginally
gravitationally-bound cloud, they show only a mild increase in cloud
size until they are eventually dispersed by radiation pressure.

\item[2.]{\it Circumcluster Surface Density Distribution}

Analytical models proposed recently by \cite{ThompsonKrumholz2016} and
in Paper I suggest that the SFE and properties of radiation-driven
outflows from star-forming regions are strongly affected by
turbulence, which establishes the distribution of gas densities.  For
clouds with centrally-concentrated stars, the relevant distributions
are the PDFs (by mass and by solid angle) as a function of circumcluster
surface (or column) density.  If most of the mass is in structures
well above the mean surface density, a GMC can simultaneously drive
low-density outflows while continuing to form stars well beyond the
point at which a uniform cloud would nominally become super-Eddington.

We find that the circumcluster distribution by mass is quite similar
to the observed column density distribution, as might be expected for
a highly filamentary cloud, and it is reasonably well-fit by a lognormal
with $\sigma_{\rm ln \Sigma} \sim 1.4$ across all clouds.  
Moreover, we
find (similar to \citealt{Dale2012, Dale2013a} for the case of ionizing
radiation) that the shape of the mass PDF is virtually unaffected by
radiation feedback up to late times.  As a result, 
much of the mass in the cloud is sub-Eddington even when the cloud has
an average Eddington factor above unity. In fact, star formation in our models
only halts when the mean value of $f_{\rm Edd} \sim 10$ is reached,
i.e., radiation pressure becomes a factor of 10 stronger than gravity (see
Figure~\ref{Fig:FEdd}).  This is substantially different from the
assumptions of most analytical models \citep[e.g.,][]{Murray2005,
  KrumholzMatzner2009, Murray2010,Fall2010,Kim2016}, which infer much
lower SFE by assuming that star formation halts once the mean
Eddington factor of $f_{\rm Edd} = 1$ is attained.

{ As a caveat, we note that our values of $\sigma_{\rm ln \Sigma}$
  may be somewhat higher than is typical in real star-forming GMCs,
  although the observational situation remains controversial.  Since
  the values of $f_{\rm Edd}$ and the SFE required to halt star
  formation increase with $\sigma_{\rm ln \Sigma}$, realistic values
  might therefore be lower than we find here.  }

\item[3.]{\it Escaping Radiation and Cloud Porosity} 

While mass PDFs are well fit by lognormals in the circumcluster
surface density $\Sigma^c$ (see e.g., Figure~ \ref{Fig:LogNormalStar}), the
solid-angle PDF is lognormal only at high $\Sigma^c$.
This implies that the optical depth to
UV radiation is less than unity over most of the sky
(e.g., Figure~\ref{Fig:Tau}), even in our highest surface density
models.  Over much of a cloud's lifetime, 50\% or more of the
radiation can escape (see e.g., Figures~\ref{Fig:FiducialAbs}, \ref{Fig:AbsCum}),
rather than contributing to the driving of gaseous outflows.

Cumulatively, the absorption fraction is $50-80~\%$ at 3 Myr 
after star formation begins (when the first SNe would explode), and $30-60~\%$
after 8 Myr (when the luminosity would drop to half its original value).
These results apply across clouds of all surface densities:
higher-surface-density clouds absorb more radiation at coeval stages
in their evolution, but they also have shorter freefall times and
therefore evolve more rapidly compared to the stellar evolution time.
Because higher-$\Sigma_{\rm cl}$ clouds have higher SFE,
clusters that form within these clouds would contribute proportionally
more to the ambient UV radiation field within galaxies.

{If $\sigma_{\ln \Sigma}$ is lower in real clouds than in our simulations,
  it would tend to reduce the escape of radiation.
  }

\item[4.]{\it Gas Outflows}
  
We find that the mean velocity (or momentum/mass) of outflowing gas in
our simulations ranges over $\sim (1.5-2.5) v_{\rm esc}(r_{\rm cl,0})$
(see Figure~\ref{Fig:VSigmaVesc}).
Figure~\ref{Fig:VInfSigma}
shows that the mean velocity of the outflow increases secularly with cloud
mass.  

We also consider the relationships between velocity and circumcluster surface
density in our models, as well as the distribution of mass with velocity.  We
are able to interpret both of these relationships in terms of an essentially
``ballistic'' model for ejection of structures by radiation forces, with
the velocity-surface density relationship given by Equation~(\ref{Eq:vrf}).
Lower-surface-density regions are preferentially driven 
to higher velocity and evidently remain coherent long enough to match
the expected relationship (see Figure~\ref{Fig:FiducialVSigma}). 

The distribution of mass as a function of velocity for outflowing gas
 (Figure~\ref{Fig:FiducialV})
extends to velocities well above the escape speed $\sim v_{\rm esc}$
from the surface of the cloud.  A lognormal distribution of mass as a
function of $\Sigma^c$ at a given launch radius maps to a distribution
of mass vs. velocity at the outer edge of the simulation domain.  The
overall mass-velocity relationship can then be understood as the
superposition of distributions of material launched from a range of
radii within the cloud with a known radial density profile
(see Fig.~\ref{Fig:FiducialVSigma}).  The
high-velocity tail originates deepest in the potential well and
experiences the strongest radiation forces.

Equation~(\ref{Eq:vrasy}) shows that at large distance from the launch
point $r_0$, optically-thin structures would reach \mbox{$\sim 9 v_{\rm
  esc}(r_0)$}, while structures that are marginally Eddington would
reach $\sim 3.7 v_{\rm esc}(r_0)$. These velocities are a few to several 
tens of ${\rm km~s^{-1}}$ for the clouds and clusters we consider.
However, \citet{ThompsonKrumholz2016} suggest that a similar dynamical process
of radiative cloud acceleration may also hold in extreme
starburst systems with high $v_{\rm esc}$, in which case the line profiles 
of high-velocity cool gas could be calculated using a similar formalism to
that given here, as expressed in Equation~(\ref{Eq:Pdv}).

\end{itemize}

{Finally, we reiterate that the numerical simulations analyzed
  here are idealized in several respects, and in this sense are best
  thought of as controlled numerical experiments rather than
  comprehensive models of real clouds.  As such, this study
  provides, for the first
  time, a systematic investigation of the radiation-matter interaction
  in turbulent, self-gravitating, uniformly-cold, unmagnetized clouds
  with localized collapse and UV feedback, which can be the basis for
  future studies with more comprehensive physics.}

\acknowledgements

We are grateful to the referee for a thorough and insightful report.  
This work was supported by Grant No. AST-1312006 from the National
Science Foundation. MAS was supported by the Max Planck/Princeton Center 
for Plasma Astrophysics under grant NSF PHY-1144374.
Part of this project was conducted during a visit
to the KITP at U.C. Santa Barbara, which is supported by the National
Science Foundation under Grant No. NSF PHY11-25915. Simulations were 
performed on the computational resources supported by the PICSciE TIGRESS 
High Performance Computing Center at Princeton University.


\bibliographystyle{apj}
\bibliography{refs}

\begin{thebibliography}{130}
\expandafter\ifx\csname natexlab\endcsname\relax\def\natexlab#1{#1}\fi

\bibitem[{{Adams}(2000)}]{Adams2000}
{Adams}, F.~C. 2000, \apj, 542, 964

\bibitem[{{Andr{\'e}} {et~al.}(2014){Andr{\'e}}, {Di Francesco},
  {Ward-Thompson}, {Inutsuka}, {Pudritz}, \& {Pineda}}]{Andre2014}
{Andr{\'e}}, P., {Di Francesco}, J., {Ward-Thompson}, D., {et~al.} 2014,
  Protostars and Planets VI, 27

\bibitem[{{Ballesteros-Paredes} {et~al.}(2011){Ballesteros-Paredes},
  {V{\'a}zquez-Semadeni}, {Gazol}, {Hartmann}, {Heitsch}, \&
  {Col{\'{\i}}n}}]{BallesterosParedes2011}
{Ballesteros-Paredes}, J., {V{\'a}zquez-Semadeni}, E., {Gazol}, A., {et~al.}
  2011, \mnras, 416, 1436

\bibitem[{{Banerjee} {et~al.}(2009){Banerjee}, {V{\'a}zquez-Semadeni},
  {Hennebelle}, \& {Klessen}}]{Banerjee2009}
{Banerjee}, R., {V{\'a}zquez-Semadeni}, E., {Hennebelle}, P., \& {Klessen},
  R.~S. 2009, \mnras, 398, 1082

\bibitem[{{Baumgardt} \& {Kroupa}(2007)}]{BaumgardtKroupa2007}
{Baumgardt}, H., \& {Kroupa}, P. 2007, \mnras, 380, 1589

\bibitem[{{Boily} \& {Kroupa}(2003)}]{BoilyKroupa2003}
{Boily}, C.~M., \& {Kroupa}, P. 2003, \mnras, 338, 665

\bibitem[{{Brunt} {et~al.}(2010){Brunt}, {Federrath}, \& {Price}}]{Brunt2010}
{Brunt}, C.~M., {Federrath}, C., \& {Price}, D.~J. 2010, \mnras, 403, 1507

\bibitem[{{Bunker} {et~al.}(2010){Bunker}, {Wilkins}, {Ellis}, {Stark},
  {Lorenzoni}, {Chiu}, {Lacy}, {Jarvis}, \& {Hickey}}]{Bunker2010}
{Bunker}, A.~J., {Wilkins}, S., {Ellis}, R.~S., {et~al.} 2010, \mnras, 409, 855

\bibitem[{{Butler} {et~al.}(2014){Butler}, {Tan}, \&
  {Kainulainen}}]{Butler2014}
{Butler}, M.~J., {Tan}, J.~C., \& {Kainulainen}, J. 2014, \apjl, 782, L30

\bibitem[{{Chevalier} \& {Clegg}(1985)}]{ChevalierClegg1985}
{Chevalier}, R.~A., \& {Clegg}, A.~W. 1985, \nat, 317, 44

\bibitem[{{Col{\'{\i}}n} {et~al.}(2013){Col{\'{\i}}n}, {V{\'a}zquez-Semadeni},
  \& {G{\'o}mez}}]{Colin2013}
{Col{\'{\i}}n}, P., {V{\'a}zquez-Semadeni}, E., \& {G{\'o}mez}, G.~C. 2013,
  \mnras, 435, 1701

\bibitem[{{Collins} {et~al.}(2012){Collins}, {Kritsuk}, {Padoan}, {Li}, {Xu},
  {Ustyugov}, \& {Norman}}]{Collins2012}
{Collins}, D.~C., {Kritsuk}, A.~G., {Padoan}, P., {et~al.} 2012, \apj, 750, 13

\bibitem[{{Da Rio} {et~al.}(2014){Da Rio}, {Tan}, \& {Jaehnig}}]{DaRio2014}
{Da Rio}, N., {Tan}, J.~C., \& {Jaehnig}, K. 2014, \apj, 795, 55

\bibitem[{{Dale} {et~al.}(2005){Dale}, {Bonnell}, {Clarke}, \&
  {Bate}}]{Dale2005}
{Dale}, J.~E., {Bonnell}, I.~A., {Clarke}, C.~J., \& {Bate}, M.~R. 2005,
  \mnras, 358, 291

\bibitem[{{Dale} {et~al.}(2012){Dale}, {Ercolano}, \& {Bonnell}}]{Dale2012}
{Dale}, J.~E., {Ercolano}, B., \& {Bonnell}, I.~A. 2012, \mnras, 424, 377

\bibitem[{{Dale} {et~al.}(2013){Dale}, {Ercolano}, \& {Bonnell}}]{Dale2013a}
---. 2013, \mnras, 430, 234

\bibitem[{{Davis} {et~al.}(2014){Davis}, {Jiang}, {Stone}, \&
  {Murray}}]{Davis2014}
{Davis}, S.~W., {Jiang}, Y.-F., {Stone}, J.~M., \& {Murray}, N. 2014, \apj,
  796, 107

\bibitem[{{Dekel} \& {Krumholz}(2013)}]{DekelKrumholz2013}
{Dekel}, A., \& {Krumholz}, M.~R. 2013, \mnras, 432, 455

\bibitem[{Dobbs {et~al.}(2013)Dobbs, Krumholz, Ballesteros-Paredes, Bolatto,
  Fukui, Heyer, Mac~Low, Ostriker, \& V{\'a}zquez-Semadeni}]{Dobbs2013}
Dobbs, C.~L., Krumholz, M.~R., Ballesteros-Paredes, J., {et~al.} 2013,
  arXiv.org

\bibitem[{{Dopita} {et~al.}(2006){Dopita}, {Fischera}, {Sutherland}, {Kewley},
  {Tuffs}, {Popescu}, {van Breugel}, {Groves}, \& {Leitherer}}]{Dopita2006}
{Dopita}, M.~A., {Fischera}, J., {Sutherland}, R.~S., {et~al.} 2006, \apj, 647,
  244

\bibitem[{{Draine}(2011)}]{Draine2011}
{Draine}, B.~T. 2011, \apj, 732, 100

\bibitem[{{Elmegreen}(1983)}]{Elmegreen1983}
{Elmegreen}, B.~G. 1983, \mnras, 203, 1011

\bibitem[{{Elmegreen} \& {Scalo}(2004)}]{ElmegreenScalo2004}
{Elmegreen}, B.~G., \& {Scalo}, J. 2004, \araa, 42, 211

\bibitem[{{Fall} {et~al.}(2010){Fall}, {Krumholz}, \& {Matzner}}]{Fall2010}
{Fall}, S.~M., {Krumholz}, M.~R., \& {Matzner}, C.~D. 2010, \apjl, 710, L142

\bibitem[{{Faucher-Gigu{\`e}re} {et~al.}(2008){Faucher-Gigu{\`e}re}, {Lidz},
  {Hernquist}, \& {Zaldarriaga}}]{Faucher-Giguere2008}
{Faucher-Gigu{\`e}re}, C.-A., {Lidz}, A., {Hernquist}, L., \& {Zaldarriaga}, M.
  2008, \apj, 688, 85

\bibitem[{{Federrath} {et~al.}(2010){Federrath}, {Banerjee}, {Clark}, \&
  {Klessen}}]{Federrath2010}
{Federrath}, C., {Banerjee}, R., {Clark}, P.~C., \& {Klessen}, R.~S. 2010,
  \apj, 713, 269

\bibitem[{{Federrath} \& {Klessen}(2013)}]{FederrathKlessen2013}
{Federrath}, C., \& {Klessen}, R.~S. 2013, \apj, 763, 51

\bibitem[{{Federrath} {et~al.}(2008){Federrath}, {Klessen}, \&
  {Schmidt}}]{Federrath2008}
{Federrath}, C., {Klessen}, R.~S., \& {Schmidt}, W. 2008, \apjl, 688, L79

\bibitem[{{Franx} {et~al.}(1997){Franx}, {Illingworth}, {Kelson}, {van Dokkum},
  \& {Tran}}]{Franx1997}
{Franx}, M., {Illingworth}, G.~D., {Kelson}, D.~D., {van Dokkum}, P.~G., \&
  {Tran}, K.-V. 1997, \apjl, 486, L75

\bibitem[{{Geyer} \& {Burkert}(2001)}]{GeyerBurkert2001}
{Geyer}, M.~P., \& {Burkert}, A. 2001, \mnras, 323, 988

\bibitem[{{Gnedin} \& {Abel}(2001)}]{GnedinAbel2001}
{Gnedin}, N.~Y., \& {Abel}, T. 2001, \na, 6, 437

\bibitem[{{Goldbaum} {et~al.}(2011){Goldbaum}, {Krumholz}, {Matzner}, \&
  {McKee}}]{Goldbaum2011}
{Goldbaum}, N.~J., {Krumholz}, M.~R., {Matzner}, C.~D., \& {McKee}, C.~F. 2011,
  \apj, 738, 101

\bibitem[{{Gong} \& {Ostriker}(2013)}]{GongOstriker2013}
{Gong}, H., \& {Ostriker}, E.~C. 2013, \apjs, 204, 8

\bibitem[{{Goodman} {et~al.}(2009){Goodman}, {Pineda}, \&
  {Schnee}}]{Goodman2009}
{Goodman}, A.~A., {Pineda}, J.~E., \& {Schnee}, S.~L. 2009, \apj, 692, 91

\bibitem[{{Goodwin}(1997)}]{Goodwin1997}
{Goodwin}, S.~P. 1997, \mnras, 284, 785

\bibitem[{{Goodwin} \& {Bastian}(2006)}]{GoodwinBastian2006}
{Goodwin}, S.~P., \& {Bastian}, N. 2006, \mnras, 373, 752

\bibitem[{{Harwit}(1962)}]{Harwit1962}
{Harwit}, M. 1962, \apj, 136, 832

\bibitem[{{Heckman} {et~al.}(1990){Heckman}, {Armus}, \& {Miley}}]{Heckman1990}
{Heckman}, T.~M., {Armus}, L., \& {Miley}, G.~K. 1990, \apjs, 74, 833

\bibitem[{{Heyer} \& {Dame}(2015)}]{HeyerDame2015}
{Heyer}, M., \& {Dame}, T.~M. 2015, \araa, 53, 583

\bibitem[{{Hockney} \& {Eastwood}(1981)}]{HockneyEastwood1981}
{Hockney}, R.~W., \& {Eastwood}, J.~W. 1981, {Computer Simulation Using
  Particles}

\bibitem[{{Hoopes} \& {Walterbos}(2000)}]{HoopesWalterbos2000}
{Hoopes}, C.~G., \& {Walterbos}, R.~A.~M. 2000, \apj, 541, 597

\bibitem[{{Hopkins} {et~al.}(2011){Hopkins}, {Quataert}, \&
  {Murray}}]{Hopkins2011}
{Hopkins}, P.~F., {Quataert}, E., \& {Murray}, N. 2011, \mnras, 417, 950

\bibitem[{{Hopkins} {et~al.}(2012){Hopkins}, {Quataert}, \&
  {Murray}}]{Hopkins2012}
---. 2012, \mnras, 421, 3488

\bibitem[{{Iffrig} \& {Hennebelle}(2015)}]{IffrigHennebelle2015}
{Iffrig}, O., \& {Hennebelle}, P. 2015, \aap, 576, A95

\bibitem[{{Kainulainen} {et~al.}(2009){Kainulainen}, {Beuther}, {Henning}, \&
  {Plume}}]{Kainulainen2009}
{Kainulainen}, J., {Beuther}, H., {Henning}, T., \& {Plume}, R. 2009, \aap,
  508, L35

\bibitem[{{Kim} \& {Ostriker}(2015)}]{KimOstriker2015}
{Kim}, C.-G., \& {Ostriker}, E.~C. 2015, \apj, 802, 99

\bibitem[{{Kim} {et~al.}(2016){Kim}, {Kim}, \& {Ostriker}}]{Kim2016}
{Kim}, J.-G., {Kim}, W.-T., \& {Ostriker}, E.~C. 2016, ArXiv e-prints

\bibitem[{{Klessen} {et~al.}(2000){Klessen}, {Heitsch}, \& {Mac
  Low}}]{Klessen2000}
{Klessen}, R.~S., {Heitsch}, F., \& {Mac Low}, M.-M. 2000, \apj, 535, 887

\bibitem[{{Kritsuk} {et~al.}(2011){Kritsuk}, {Norman}, \&
  {Wagner}}]{Kritsuk2011}
{Kritsuk}, A.~G., {Norman}, M.~L., \& {Wagner}, R. 2011, \apjl, 727, L20

\bibitem[{{Krumholz} \& {Dekel}(2010)}]{KrumholzDekel2010}
{Krumholz}, M.~R., \& {Dekel}, A. 2010, \mnras, 406, 112

\bibitem[{{Krumholz} {et~al.}(2007){Krumholz}, {Klein}, {McKee}, \&
  {Bolstad}}]{Krumholz2007}
{Krumholz}, M.~R., {Klein}, R.~I., {McKee}, C.~F., \& {Bolstad}, J. 2007, \apj,
  667, 626

\bibitem[{{Krumholz} \& {Matzner}(2009)}]{KrumholzMatzner2009}
{Krumholz}, M.~R., \& {Matzner}, C.~D. 2009, \apj, 703, 1352

\bibitem[{{Krumholz} \& {Thompson}(2012)}]{KrumholzThompson2012}
{Krumholz}, M.~R., \& {Thompson}, T.~A. 2012, \apj, 760, 155

\bibitem[{{Krumholz} \& {Thompson}(2013)}]{KrumholzThompson2013}
---. 2013, \mnras, 434, 2329

\bibitem[{{Kuhlen} \& {Faucher-Gigu{\`e}re}(2012)}]{KuhlenFaucher-Giguere2012}
{Kuhlen}, M., \& {Faucher-Gigu{\`e}re}, C.-A. 2012, \mnras, 423, 862

\bibitem[{{Lada} \& {Lada}(2003)}]{LadaLada2003}
{Lada}, C.~J., \& {Lada}, E.~A. 2003, \araa, 41, 57

\bibitem[{{Lada} {et~al.}(1984){Lada}, {Margulis}, \& {Dearborn}}]{Lada1984}
{Lada}, C.~J., {Margulis}, M., \& {Dearborn}, D. 1984, \apj, 285, 141

\bibitem[{{Larson}(1969)}]{Larson1969}
{Larson}, R.~B. 1969, \mnras, 145, 271

\bibitem[{{Lee} {et~al.}(2015){Lee}, {Chang}, \& {Murray}}]{Lee2014}
{Lee}, E.~J., {Chang}, P., \& {Murray}, N. 2015, \apj, 800, 49

\bibitem[{{Levermore}(1984)}]{Levermore1984}
{Levermore}, C.~D. 1984, \jqsrt, 31, 149

\bibitem[{{Lim} {et~al.}(2016){Lim}, {Tan}, {Kainulainen}, {Ma}, \&
  {Butler}}]{Lim2016}
{Lim}, W., {Tan}, J.~C., {Kainulainen}, J., {Ma}, B., \& {Butler}, M.~J. 2016,
  \apjl, 829, L19

\bibitem[{{Lombardi} {et~al.}(2015){Lombardi}, {Alves}, \&
  {Lada}}]{Lombardi2015}
{Lombardi}, M., {Alves}, J., \& {Lada}, C.~J. 2015, \aap, 576, L1

\bibitem[{{Lombardi} {et~al.}(2010){Lombardi}, {Lada}, \&
  {Alves}}]{Lombardi2010}
{Lombardi}, M., {Lada}, C.~J., \& {Alves}, J. 2010, \aap, 512, A67

\bibitem[{{Longmore} {et~al.}(2014){Longmore}, {Kruijssen}, {Bastian}, {Bally},
  {Rathborne}, {Testi}, {Stolte}, {Dale}, {Bressert}, \&
  {Alves}}]{Longmore2014}
{Longmore}, S.~N., {Kruijssen}, J.~M.~D., {Bastian}, N., {et~al.} 2014,
  Protostars and Planets VI, 291

\bibitem[{{Lopez} {et~al.}(2011){Lopez}, {Krumholz}, {Bolatto}, {Prochaska}, \&
  {Ramirez-Ruiz}}]{Lopez2011}
{Lopez}, L.~A., {Krumholz}, M.~R., {Bolatto}, A.~D., {Prochaska}, J.~X., \&
  {Ramirez-Ruiz}, E. 2011, \apj, 731, 91

\bibitem[{{Lopez} {et~al.}(2014){Lopez}, {Krumholz}, {Bolatto}, {Prochaska},
  {Ramirez-Ruiz}, \& {Castro}}]{Lopez2014}
{Lopez}, L.~A., {Krumholz}, M.~R., {Bolatto}, A.~D., {et~al.} 2014, \apj, 795,
  121

\bibitem[{{Mac Low} \& {Klessen}(2004)}]{MacLowKlessen2004}
{Mac Low}, M.-M., \& {Klessen}, R.~S. 2004, Reviews of Modern Physics, 76, 125

\bibitem[{{Madau} {et~al.}(1999){Madau}, {Haardt}, \& {Rees}}]{Madau1999}
{Madau}, P., {Haardt}, F., \& {Rees}, M.~J. 1999, \apj, 514, 648

\bibitem[{{Martin}(2005)}]{Martin2005}
{Martin}, C.~L. 2005, \apj, 621, 227

\bibitem[{{Martizzi} {et~al.}(2015){Martizzi}, {Faucher-Gigu{\`e}re}, \&
  {Quataert}}]{Martizzi2015}
{Martizzi}, D., {Faucher-Gigu{\`e}re}, C.-A., \& {Quataert}, E. 2015, \mnras,
  450, 504

\bibitem[{{Matzner}(2002)}]{Matzner2002}
{Matzner}, C.~D. 2002, \apj, 566, 302

\bibitem[{{Matzner} \& {Jumper}(2015)}]{MatznerJumper2015}
{Matzner}, C.~D., \& {Jumper}, P.~H. 2015, \apj, 815, 68

\bibitem[{{McKee} \& {Ostriker}(2007)}]{McKeeOstriker2007}
{McKee}, C.~F., \& {Ostriker}, E.~C. 2007, \araa, 45, 565

\bibitem[{{McKee} \& {Tan}(2003)}]{MckeeTan2003}
{McKee}, C.~F., \& {Tan}, J.~C. 2003, \apj, 585, 850

\bibitem[{{M{\'e}nard} {et~al.}(2011){M{\'e}nard}, {Wild}, {Nestor}, {Quider},
  {Zibetti}, {Rao}, \& {Turnshek}}]{Menard2009}
{M{\'e}nard}, B., {Wild}, V., {Nestor}, D., {et~al.} 2011, \mnras, 417, 801

\bibitem[{{Murray}(2011)}]{Murray2011}
{Murray}, N. 2011, \apj, 729, 133

\bibitem[{{Murray} {et~al.}(2005){Murray}, {Quataert}, \&
  {Thompson}}]{Murray2005}
{Murray}, N., {Quataert}, E., \& {Thompson}, T.~A. 2005, \apj, 618, 569

\bibitem[{{Murray} {et~al.}(2010){Murray}, {Quataert}, \&
  {Thompson}}]{Murray2010}
---. 2010, \apj, 709, 191

\bibitem[{{O'Dell} {et~al.}(1967){O'Dell}, {York}, \& {Henize}}]{Odell1967}
{O'Dell}, C.~R., {York}, D.~G., \& {Henize}, K.~G. 1967, \apj, 150, 835

\bibitem[{Olsen {et~al.}(2015)Olsen, Greve, Brinch, Sommer-Larsen, Rasmussen,
  Toft, \& Zirm}]{Olsen2015}
Olsen, K.~P., Greve, T.~R., Brinch, C., {et~al.} 2015, arXiv.org

\bibitem[{{Ostriker} \& {Shetty}(2011)}]{OstrikerShetty2011}
{Ostriker}, E.~C., \& {Shetty}, R. 2011, \apj, 731, 41

\bibitem[{{Ostriker} {et~al.}(2001){Ostriker}, {Stone}, \&
  {Gammie}}]{Ostriker2001}
{Ostriker}, E.~C., {Stone}, J.~M., \& {Gammie}, C.~F. 2001, \apj, 546, 980

\bibitem[{{Padoan} {et~al.}(2004{\natexlab{a}}){Padoan}, {Jimenez}, {Juvela},
  \& {Nordlund}}]{Padoan2004b}
{Padoan}, P., {Jimenez}, R., {Juvela}, M., \& {Nordlund}, {\AA}.
  2004{\natexlab{a}}, \apjl, 604, L49

\bibitem[{{Padoan} {et~al.}(2004{\natexlab{b}}){Padoan}, {Jimenez}, {Nordlund},
  \& {Boldyrev}}]{Padoan2004a}
{Padoan}, P., {Jimenez}, R., {Nordlund}, {\AA}., \& {Boldyrev}, S.
  2004{\natexlab{b}}, Physical Review Letters, 92, 191102

\bibitem[{{Pellegrini} {et~al.}(2010){Pellegrini}, {Baldwin}, \&
  {Ferland}}]{Pellegrini2010}
{Pellegrini}, E.~W., {Baldwin}, J.~A., \& {Ferland}, G.~J. 2010, \apjs, 191,
  160

\bibitem[{{Pellegrini} {et~al.}(2007){Pellegrini}, {Baldwin}, {Brogan},
  {Hanson}, {Abel}, {Ferland}, {Nemala}, {Shaw}, \& {Troland}}]{Pellegrini2007}
{Pellegrini}, E.~W., {Baldwin}, J.~A., {Brogan}, C.~L., {et~al.} 2007, \apj,
  658, 1119

\bibitem[{{Penston}(1969)}]{Penston1969}
{Penston}, M.~V. 1969, \mnras, 144, 425

\bibitem[{{Pettini} {et~al.}(2001){Pettini}, {Shapley}, {Steidel}, {Cuby},
  {Dickinson}, {Moorwood}, {Adelberger}, \& {Giavalisco}}]{Pettini2001}
{Pettini}, M., {Shapley}, A.~E., {Steidel}, C.~C., {et~al.} 2001, \apj, 554,
  981

\bibitem[{{Pettini} {et~al.}(2000){Pettini}, {Steidel}, {Adelberger},
  {Dickinson}, \& {Giavalisco}}]{Pettini2000}
{Pettini}, M., {Steidel}, C.~C., {Adelberger}, K.~L., {Dickinson}, M., \&
  {Giavalisco}, M. 2000, \apj, 528, 96

\bibitem[{{Pfalzner} {et~al.}(2016){Pfalzner}, {Kirk}, {Sills}, {Urquhart},
  {Kauffmann}, {Kuhn}, {Bhandare}, \& {Menten}}]{Pfalzner2016}
{Pfalzner}, S., {Kirk}, H., {Sills}, A., {et~al.} 2016, \aap, 586, A68

\bibitem[{{Plummer}(1911)}]{Plummer1911}
{Plummer}, H.~C. 1911, \mnras, 71, 460

\bibitem[{{Price} {et~al.}(2011){Price}, {Federrath}, \& {Brunt}}]{Price2011}
{Price}, D.~J., {Federrath}, C., \& {Brunt}, C.~M. 2011, \apjl, 727, L21

\bibitem[{{Proszkow} \& {Adams}(2009)}]{ProszkowAdams2009}
{Proszkow}, E.-M., \& {Adams}, F.~C. 2009, \apjs, 185, 486

\bibitem[{{Raskutti} {et~al.}(2016){Raskutti}, {Ostriker}, \&
  {Skinner}}]{Raskutti2016}
{Raskutti}, S., {Ostriker}, E.~C., \& {Skinner}, M.~A. 2016, \apj, 829, 130

\bibitem[{{Roman-Duval} {et~al.}(2010){Roman-Duval}, {Jackson}, {Heyer},
  {Rathborne}, \& {Simon}}]{Roman-Duval2010}
{Roman-Duval}, J., {Jackson}, J.~M., {Heyer}, M., {Rathborne}, J., \& {Simon},
  R. 2010, \apj, 723, 492

\bibitem[{{Rupke} {et~al.}(2005){Rupke}, {Veilleux}, \& {Sanders}}]{Rupke2005}
{Rupke}, D.~S., {Veilleux}, S., \& {Sanders}, D.~B. 2005, \apjs, 160, 115

\bibitem[{{Sales} {et~al.}(2014){Sales}, {Marinacci}, {Springel}, \&
  {Petkova}}]{Sales2014}
{Sales}, L.~V., {Marinacci}, F., {Springel}, V., \& {Petkova}, M. 2014, \mnras,
  439, 2990

\bibitem[{{Scannapieco} \& {Br{\"u}ggen}(2015)}]{2015ApJ...805..158S}
{Scannapieco}, E., \& {Br{\"u}ggen}, M. 2015, \apj, 805, 158

\bibitem[{{Schneider} {et~al.}(2013){Schneider}, {Andr{\'e}}, {K{\"o}nyves},
  {Bontemps}, {Motte}, {Federrath}, {Ward-Thompson}, {Arzoumanian},
  {Benedettini}, {Bressert}, {Didelon}, {Di Francesco}, {Griffin}, {Hennemann},
  {Hill}, {Palmeirim}, {Pezzuto}, {Peretto}, {Roy}, {Rygl}, {Spinoglio}, \&
  {White}}]{Schneider2013}
{Schneider}, N., {Andr{\'e}}, P., {K{\"o}nyves}, V., {et~al.} 2013, \apjl, 766,
  L17

\bibitem[{{Schneider} {et~al.}(2015){Schneider}, {Ossenkopf}, {Csengeri},
  {Klessen}, {Federrath}, {Tremblin}, {Girichidis}, {Bontemps}, \&
  {Andr{\'e}}}]{Schneider2015}
{Schneider}, N., {Ossenkopf}, V., {Csengeri}, T., {et~al.} 2015, \aap, 575, A79

\bibitem[{{Scoville}(2003)}]{Scoville2003}
{Scoville}, N. 2003, Journal of Korean Astronomical Society, 36, 167

\bibitem[{{Scoville} {et~al.}(2001){Scoville}, {Polletta}, {Ewald}, {Stolovy},
  {Thompson}, \& {Rieke}}]{Scoville2001}
{Scoville}, N.~Z., {Polletta}, M., {Ewald}, S., {et~al.} 2001, \aj, 122, 3017

\bibitem[{{Scoville} \& {Solomon}(1975)}]{ScovilleSolomon1975}
{Scoville}, N.~Z., \& {Solomon}, P.~M. 1975, \apjl, 199, L105

\bibitem[{{Scoville} {et~al.}(1987){Scoville}, {Yun}, {Sanders}, {Clemens}, \&
  {Waller}}]{Scoville1987}
{Scoville}, N.~Z., {Yun}, M.~S., {Sanders}, D.~B., {Clemens}, D.~P., \&
  {Waller}, W.~H. 1987, \apjs, 63, 821

\bibitem[{{Shapley} {et~al.}(2003){Shapley}, {Steidel}, {Pettini}, \&
  {Adelberger}}]{Shapley2003}
{Shapley}, A.~E., {Steidel}, C.~C., {Pettini}, M., \& {Adelberger}, K.~L. 2003,
  \apj, 588, 65

\bibitem[{{Skinner} \& {Ostriker}(2013)}]{SkinnerOstriker2013}
{Skinner}, M.~A., \& {Ostriker}, E.~C. 2013, \apjs, 206, 21

\bibitem[{{Skinner} \& {Ostriker}(2015)}]{SkinnerOstriker2015}
---. 2015, \apj, 809, 187

\bibitem[{{Springel}(2005)}]{Springel2005}
{Springel}, V. 2005, \mnras, 364, 1105

\bibitem[{{Steidel} {et~al.}(1996){Steidel}, {Giavalisco}, {Pettini},
  {Dickinson}, \& {Adelberger}}]{Steidel1996}
{Steidel}, C.~C., {Giavalisco}, M., {Pettini}, M., {Dickinson}, M., \&
  {Adelberger}, K.~L. 1996, \apjl, 462, L17

\bibitem[{{Stone} \& {Gardiner}(2009)}]{StoneGardiner2009}
{Stone}, J.~M., \& {Gardiner}, T. 2009, \na, 14, 139

\bibitem[{{Stone} {et~al.}(2008){Stone}, {Gardiner}, {Teuben}, {Hawley}, \&
  {Simon}}]{Stone2008}
{Stone}, J.~M., {Gardiner}, T.~A., {Teuben}, P., {Hawley}, J.~F., \& {Simon},
  J.~B. 2008, \apjs, 178, 137

\bibitem[{{Stone} {et~al.}(1998){Stone}, {Ostriker}, \& {Gammie}}]{Stone1998}
{Stone}, J.~M., {Ostriker}, E.~C., \& {Gammie}, C.~F. 1998, \apjl, 508, L99

\bibitem[{{Thompson} {et~al.}(2015){Thompson}, {Fabian}, {Quataert}, \&
  {Murray}}]{Thompson2015}
{Thompson}, T.~A., {Fabian}, A.~C., {Quataert}, E., \& {Murray}, N. 2015,
  \mnras, 449, 147

\bibitem[{{Thompson} \& {Krumholz}(2016)}]{ThompsonKrumholz2016}
{Thompson}, T.~A., \& {Krumholz}, M.~R. 2016, \mnras, 455, 334

\bibitem[{{Thompson} {et~al.}(2005){Thompson}, {Quataert}, \&
  {Murray}}]{Thompson2005}
{Thompson}, T.~A., {Quataert}, E., \& {Murray}, N. 2005, \apj, 630, 167

\bibitem[{{Thompson} {et~al.}(2016){Thompson}, {Quataert}, {Zhang}, \&
  {Weinberg}}]{Thompson2016}
{Thompson}, T.~A., {Quataert}, E., {Zhang}, D., \& {Weinberg}, D.~H. 2016,
  \mnras, 455, 1830

\bibitem[{{Tremonti} {et~al.}(2007){Tremonti}, {Moustakas}, \&
  {Diamond-Stanic}}]{Tremonti2007}
{Tremonti}, C.~A., {Moustakas}, J., \& {Diamond-Stanic}, A.~M. 2007, \apjl,
  663, L77

\bibitem[{{V{\'a}zquez-Semadeni} {et~al.}(2011){V{\'a}zquez-Semadeni},
  {Banerjee}, {G{\'o}mez}, {Hennebelle}, {Duffin}, \&
  {Klessen}}]{Vazquez-Semadeni2011}
{V{\'a}zquez-Semadeni}, E., {Banerjee}, R., {G{\'o}mez}, G.~C., {et~al.} 2011,
  \mnras, 414, 2511

\bibitem[{{V{\'a}zquez-Semadeni} {et~al.}(2010){V{\'a}zquez-Semadeni},
  {Col{\'{\i}}n}, {G{\'o}mez}, {Ballesteros-Paredes}, \&
  {Watson}}]{Vazquez-Semadeni2010}
{V{\'a}zquez-Semadeni}, E., {Col{\'{\i}}n}, P., {G{\'o}mez}, G.~C.,
  {Ballesteros-Paredes}, J., \& {Watson}, A.~W. 2010, \apj, 715, 1302

\bibitem[{{V{\'a}zquez-Semadeni} \&
  {Garc{\'{\i}}a}(2001)}]{VazquezSemadeni2001}
{V{\'a}zquez-Semadeni}, E., \& {Garc{\'{\i}}a}, N. 2001, \apj, 557, 727

\bibitem[{{V{\'a}zquez-Semadeni} {et~al.}(2008){V{\'a}zquez-Semadeni},
  {Gonz{\'a}lez}, {Ballesteros-Paredes}, {Gazol}, \&
  {Kim}}]{Vazquez-Semadeni2008}
{V{\'a}zquez-Semadeni}, E., {Gonz{\'a}lez}, R.~F., {Ballesteros-Paredes}, J.,
  {Gazol}, A., \& {Kim}, J. 2008, \mnras, 390, 769

\bibitem[{{Veilleux} {et~al.}(2005){Veilleux}, {Cecil}, \&
  {Bland-Hawthorn}}]{Veilleux2005}
{Veilleux}, S., {Cecil}, G., \& {Bland-Hawthorn}, J. 2005, \araa, 43, 769

\bibitem[{{Voges} \& {Walterbos}(2006)}]{VogesWalterbos2006}
{Voges}, E.~S., \& {Walterbos}, R.~A.~M. 2006, \apjl, 644, L29

\bibitem[{{Walch} \& {Naab}(2015)}]{WalchNaab2015}
{Walch}, S., \& {Naab}, T. 2015, \mnras, 451, 2757

\bibitem[{{Walch} {et~al.}(2012){Walch}, {Whitworth}, {Bisbas}, {W{\"u}nsch},
  \& {Hubber}}]{Walch2012}
{Walch}, S.~K., {Whitworth}, A.~P., {Bisbas}, T., {W{\"u}nsch}, R., \&
  {Hubber}, D. 2012, \mnras, 427, 625

\bibitem[{{Wang}(1995)}]{Wang1995}
{Wang}, B. 1995, \apj, 444, 590

\bibitem[{{Weiner} {et~al.}(2009){Weiner}, {Coil}, {Prochaska}, {Newman},
  {Cooper}, {Bundy}, {Conselice}, {Dutton}, {Faber}, {Koo}, {Lotz}, {Rieke}, \&
  {Rubin}}]{Weiner2009}
{Weiner}, B.~J., {Coil}, A.~L., {Prochaska}, J.~X., {et~al.} 2009, \apj, 692,
  187

\bibitem[{{Weingartner} \& {Draine}(2001)}]{WeingartnerDraine2001}
{Weingartner}, J.~C., \& {Draine}, B.~T. 2001, \apj, 548, 296

\bibitem[{{Whitworth} \& {Ward-Thompson}(2001)}]{WhitworthWard-Thompson2001}
{Whitworth}, A.~P., \& {Ward-Thompson}, D. 2001, \apj, 547, 317

\bibitem[{{Zhang} \& {Thompson}(2012)}]{ZhangThompson2012}
{Zhang}, D., \& {Thompson}, T.~A. 2012, \mnras, 424, 1170

\end{thebibliography}


\end{document}